\documentclass{ieeeaccess}
\ifCLASSINFOpdf
\else
\fi
\usepackage{float} 
   \usepackage{caption}
	\usepackage{subcaption}
	\usepackage{graphicx, booktabs}
    \usepackage{graphics}
	\usepackage{amsmath}
	\usepackage{pdfpages}
    \usepackage{mathtools}
    \usepackage{setspace}
    \usepackage{rotating}    
\usepackage{chngcntr}
\usepackage{etoolbox}

\usepackage{amsmath}
\usepackage{multirow}
\usepackage{makecell}

\begin{document}
\history{This work has been submitted to the IEEE for possible publication. Copyright may be transferred without notice, after which this version may no longer be accessible}
\doi{TBC}
\title{Symbol-Level Synchronisation Channel Modelling With Real-World Application: From Davey-Mackay, Fritchman to Markov}


\author{\uppercase{Shamin~Achari}\authorrefmark{1},
\uppercase{Daniel~G.~Holmes\authorrefmark{1}, and Ling~Cheng}.\authorrefmark{1},
\IEEEmembership{Senior Member, IEEE}}
\address[1]{School of Electrical and Information Engineering, University of the Witwatersrand, Johannesburg, South Africa}

\tfootnote{This work was supported in part by the South African National Research Foundation, Grant Number: 129311}

\markboth
{Author \headeretal: Preparation of Papers for IEEE TRANSACTIONS and JOURNALS}
{Author \headeretal: Preparation of Papers for IEEE TRANSACTIONS and JOURNALS}

\corresp{Corresponding author: Ling~Cheng (e-mail: Ling.Cheng@wits.ac.za).}

\begin{abstract}
Errors in realistic channels contain not only substitution errors, but synchronisation errors as well. Moreover, these errors are rarely statistically independent in nature. By extending on the idea of the Fritchman channel model, a novel error category-based methodology in determining channel characteristics is described for memory channels which contain insertion, deletion, and substitution errors.The practicality of such a methodology is reinforced by making use of real communication data from a visible light communication system. Simulation results show that the error-free and error runs using this new method of defining the channel clearly deviates from the Davey-Mackay synchronisation model which is memoryless in nature. This further emphasises the inherent memory in these synchronisation channels which we are now able to characterise.  Additionally, a new method to determine the parameters of a synchronisation memory channel using the Levenshtein distance metric is detailed. This method of channel modelling allows for more realistic communication models to be simulated and can easily extend to other areas of research such as DNA barcoding in the medical domain. 
\end{abstract}

\begin{keywords}
Channel Modelling, Finite-State Markov Channel, Memory Models, Synchronisation Models.
\end{keywords}

\titlepgskip=-15pt

\maketitle

\section{Introduction}

\IEEEPARstart{S}{ystems} which exhibit a correlation between errors while also having synchronisation issues are quite common in practical, real-life applications. Thus, a method to characterise and model such systems proves beneficial. A few of these applications involve data transmission especially in cases where the channel is significantly harsh. Visible Light Communication (VLC) and Free Space Optical (FSO) communications are examples of such channels because they suffer drastically under the influence of interference, signal blocking, and turbulence. The applications are not only restricted to the domain of telecommunications either and can easily be extended to domains such as medicine. An example of this is described in \cite{kracht2014using} and \cite{kracht2015insertion} where Kracht and Schober modify the idea of watermark codes and synchronisation error channels described by Davey and Mackay \cite{mackay} to model and correct for errors while barcoding DNA in the process of DNA sequencing. Kracht et al describes how the system works well, but would benefit from a more complex model which incorporates memory into the channel as DNA sequencing channels are known to exhibit correlations between errors.

There are numerous channel models which take into account a combination of substitution errors as well as insertion and deletion errors (commonly referred to as synchronisation errors). The most cited of which include the Gallager channel model \cite{gallager1961sequential}, Zigangirov channel model \cite{zigangirov1969sequential} and the Davey-Mackay (DM) channel model \cite{mackay, davey2000error}. In terms of memory channels, an extensive review is conducted on the relevant error control techniques and is presented in \cite{kanal1978models}. More recently, \cite{sadeghi2008finite} provides an in-depth look at the modeling of FSMC for fading channels. There, however, appears to be little mentioned regarding synchronisation memory channels. The focus of this paper will remain on generative channel models where, in this case, the most cited include the Gilbert channel model \cite{gilbert1960capacity}, the Gilbert-Elliott (GE) channel model \cite{elliott1963estimates} and the Fritchman channel model \cite{fritchman1967binary}.

Channel models currently exist for systems with discrete synchronisation errors, and separately for those which are capable of characterising memory. However, to the knowledge of the authors, no such model or modeling technique exists where both scenarios are taken into account in a single model. This paper firstly introduces the idea of using the Fritchman Model and consequently Hidden Markov Models, which inherently contain memory to model substitution errors in addition to synchronisation errors. This is achieved by making use of various error groupings. Further to this, a novel Finite-State Makov Channel (FSMC) model which contains states for insertions, deletions, substitutions and transmission is presented. This provides a more comprehensive model for real world scenarios and the applicability of such a model is reinforced by using communication data from an actual VLC testbed.

The rest of the paper is structured as follows. The DM channel as well as the Fritchman model are further detailed in Section \ref{sec:background} along with the metrics used for analysis. This is followed by Section \ref{sec:SetupChan} where the system setup and approach is described. Section \ref{sec:AnaChan} then shows an analysis using a modified Fritchman model to characterise synchronisation errors. From this, a novel channel which consists of both synchronisation errors and memory, is formulated and discussed in Section \ref{sec:Novel}. Conclusions are drawn in Section \ref{sec:Conclusion}.

\section{Background}
\label{sec:background}

\subsection{Davey-Mackay Synchronisation Channel Model}
The Gallager, Zigangirov and DM models are all binary, discrete, and memoryless, which means they tend towards an independent and identically distributed (IID) classification. Additionally, none of these channels are able to indicate the positions of errors.  For this paper, we focus on the DM channel as it is the most comprehensive and incorporates elements from both the Gallager and Zigangirov models. In fact, Leigh shows that both the Zigangirov and DM channels are equivalent when the parameters are specifically defined \cite{leigh}. More comprehensive details regarding the Gallager channel model are found in \cite{gallager1961sequential,leigh, wang2012coding} and likewise further information regarding the Zigangirov channel is available in \cite{zigangirov1969sequential,leigh, wang2012coding}.

\begin{figure}[th]
\centering
\includegraphics[width=\columnwidth]{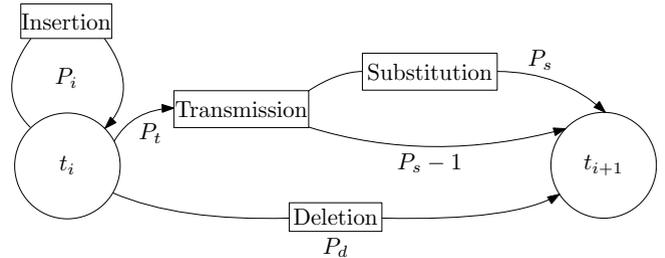}
\caption[DM synchronisation channel model]{Davey-Mackay synchronisation channel model}
\label{fig:DMChannel}
\end{figure}

In the DM model, the queued bits awaiting transmission may undergo one of three events to proceed to the next time step. Bits may be inserted into the received stream with a probability of $P_i$. Since there is, in theory, an unlimited number of possible insertions, for $n$ insertions the probability is given as $P_i^n$. A transmission or deletion must follow an insertion to allow the system to move into the next time step. A maximum number of insertions $I$ is imposed on the system for simplification. With a probability $P_d$, a bit is deleted from the stream and does not appear in the received sequence. Lastly, with a probability $P_t$, a bit is transmitted where {$P_t = 1 - P_i - P_d$} \cite{davey2000error,mackay,leigh}. Since bit flips are accounted for, a substitution error may occur on a transmitted bit with a probability $P_s$. As such, the sum of the probabilities of a substitution and no substitution must equal unity. The DM channel model is better illustrated in Figure \ref{fig:DMChannel} \cite{wang2012coding} where $t_i$  and $t_{i+1}$ indicates the time steps at time $i$ and $i+1$ respectively.

\subsection{Fritchman Channel Model and Parameter Estimation}
The simplest way of describing how errors occur in a channel is with the use of a discrete, memoryless channel, where the current output is only dependent on the current input \cite{swarts1991markov}. Given an input alphabet $X={x_0, x_1, \ldots, x_{q-1}}$ and an output alphabet  $Y={y_0, y_1, \ldots, y_{Q-1}}$, then the set of $q \times Q$ conditional probabilities are given by $P(Y=y_i|X=x_i) = P(y_i|x_i)$. These parameters are able to completely define a discrete, memoryless channel \cite{swarts1991markov, proakisdigital}. Unfortunately, most real channels exhibit some memory within the system, where the cause of one error tends to create more errors within that region of transmission \cite{swarts1991markov}. An easy method of overcoming this memory in substitution error channels is by converting them into memoryless channels with the use of interleavers, as this spreads the errors throughout the sequence making the distribution IID in nature and "locally memoryless" \cite{zhong2007binary}. While the process of interleaving simplifies the modelling it adds additional system complexity and delays while also foregoing the additional channel capacity we may gain by utilising the channels inherent memory \cite{zhong2007binary}. Additionally, the use of interleavers poses significant problems when dealing with synchronisation errors, as there is no accurate way of knowing how many and in what positions the bits were inserted and deleted. This makes it near impossible to determine the depth of the interleaver required and thus drastically restricts the use of them in these considered channels.

Since the models of interest are generative, we are able to generate statistics based on the error sequences. The alternate category of discrete channel models are descriptive and are discussed further in \cite{swarts1991markov}. The Gilbert, GE, and Fritchman models all have the ability to to model memory in a channel where the errors are binary in nature and have some statistical dependence between them \cite{swarts1991markov}. Additionally, all these models employ the use of finite-state Markov models \cite{tranter2004principles,swarts1991markov,holmes2019modelling}. This paper once again limits the discussion to Fritchman models for statistically dependent error channels as it is the most comprehensive. It has also gained substantial attention in recent years due to the practicality it offers in modelling realistic communication channels and the ease of parameter estimation \cite{tranter2004principles}. Readers may find more information of the Gilbert and GE channel models in \cite{gilbert1960capacity} and \cite{elliott1963estimates} respectively.

\begin{figure}[th]
\centering
\includegraphics[width=\columnwidth]{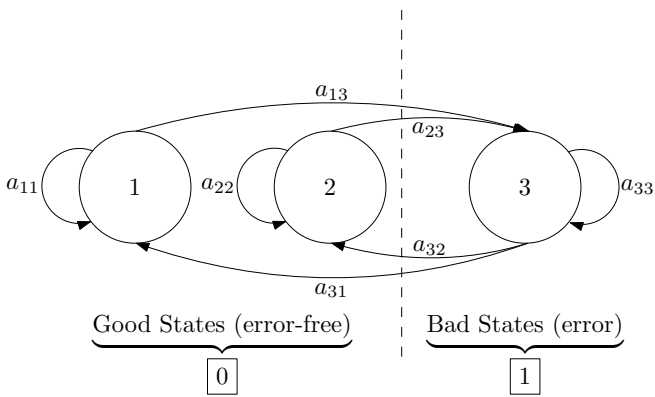}
\caption[Three state Fritchman model with single error state]{Three state Fritchman model with single error state}
\label{fig:FritchmanChannel}
\end{figure}

Fritchman used a finite state partitioned Markov model to model binary errors where the partitioning was done according to error-free and error states. The Fritchman model contains $N$ total states, of which $K$ states are partitioned as good states and the remaining $N-K$ states are bad states. A condition is also applied where a good state is error free and must produce a 0 in the error sequence, whereas a bad state is erroneous in nature and will always produce a 1 in the error sequence \cite{fritchman1967binary,swarts1991markov}. Using a single error state not only reduces complexity, but it allows the model parameters to be uniquely specified. It also reduces the model parameters from $2K(N-K)$ parameters to $2(N-1)$ parameters. In a single error state model, the error free run distribution can completely specify the model parameters \cite{fritchman1967binary}. Figure \ref{fig:FritchmanChannel} shows a simplified 3 state Fritchman model with a single error state, along with the transition and emission matrices shown in Equations \eqref{eqn:FrTrans} and \eqref{eqn:FrEm} respectively \cite{fritchman1967binary,tranter2004principles,holmes2019modelling}.

\begin{equation}
\label{eqn:FrTrans}
A = 
\begin{bmatrix} 
a_{11} & 0 & a_{13}\\
0 & a_{22} & a_{23}\\
a_{31} & a_{32} & a_{33}\\
\end{bmatrix}
\quad
\end{equation}

\begin{equation}
\label{eqn:FrEm}
B = 
\begin{bmatrix} 
1 & 1 & 0\\
0 & 0 & 1\\
\end{bmatrix}
\quad
\end{equation}

The Fritchman model makes use of an empirical technique where the exponentials from \eqref{eqn:FrEFR} and \eqref{eqn:FrER} are used to fit the measured data \cite{tranter2004principles}. Since the Fritchman model is designed on Markov processes, an easier method of parameter estimation is based on the iterative procedural Baum-Welch algorithm. This algorithm converges to a maximum likelihood estimator of $\{A,B\}$ which seeks to maximise $Pr(O|\{A,B\})$ \cite{tranter2004principles}. The Baum Welch algorithm is detailed in \cite{tranter2004principles}. It is worth noting that for the Fritchman model, only the transition matrix, $A$, is estimated as the entries for the emission matrix, $B$, are fully known. Additionally, $O$ corresponds to the observed sequence.

\subsection{Performance Metrics for Model Analysis}
\label{sec:Metrics}
To effectively analyse the proposed system, we use the Chi Squared ($\chi^2$) Goodness of fit test as well as the Mean Squared Error (MSE). These statistics will determine how well the observed data fits with the expected data. We then look at error-free and error run distributions to characterise the channel models. The $\chi^2$, MSE and procedure used for the tests are outlined in Appendix \ref{app:metricmethod} along with a brief description of the error-free and error run distributions.

\section{Error Category-Based Channel Models Setup}
\label{sec:SetupChan}

 As mentioned previously, the applications for this type of analysis is widespread. To better solidify the practicality of this methodology, data collected from a VLC testbed is used in this approach. The data is publicly available from \cite{github} and is originally used in \cite{achari2021selfsynchronising}, which describes an inter and intra-vehicle data communication system based on VLC. We limit the parameters of the VLC data to use 1 synchronisation word and a frame length of 10003 symbols to simplify the analysis, but this procedure can easily be extended to other parameters and data sets. In particular we look at low SNR, low baudrate communication, as well as high SNR, high baudrate communication as this is where the most errors in the system, without complete failure, are observed. This procedure focuses on the symbol level (bit level for binary systems) as the transitions within a synchronised frame are analysed.
 
 In this approach we assume that the receiver has full knowledge of the transmitted data and as a result, the Levenshtein Distance (edit distance) algorithm may be used to determine the most likely states (insertion, deletion, substitution or transmission) that the channel traverses during communication \cite{levenshtein1966binary,ldpeople,editDisalg}. This state path is hereon referred to as the synchronisation error sequence and it is obtained for various communication parameters. Additionally, the error probabilities for $P_t$, $P_s$, $P_i$ and $P_d$ are calculated from this synchronisation error sequence by summing the number of occurrences of a certain state and dividing this by the total length of the sequence. These error probabilities allow us to simulate communication over the DM synchronisation channel for comparison. Depending on the type of error category of interest, we can also calculate the overall error probability, $P_e$, which will be used to simulate the IID plots.
 
 An example synchronisation error sequence for an Insertion, Deletion and Substitution (IDS) channel could be, \textbf{t,t,s,t,t,t,t,i,t,t,t,t,d,t,t,t,t,t,s,t} where $t$ describes an error free transmission, an $s$ represents a bit flip or substitution error and an $i$ and $d$ represents an insertion and deletion respectively.

The next step is to generate the channel model. For simplicity, we limit the Fritchman model to three states, with a single error state, as there tends to only be slight accuracy gains with much more complexity for the additional states \cite{holmessemi}. Using this approach will require some manipulation as the model is binary in nature, and a synchronisation channel produces a variety of errors (insertions, deletions and substitutions). As such, the procedure will convert the synchronisation channel errors encountered in an error sequence into a binary error sequence. This will be used to create a channel model based on the Fritchman model and Baum-Welch algorithm to ultimately determine the parameters of the channel.  

To convert this synchronisation error sequence to a binary form, we will look at five different categorisations or error category-based groupings: Error or Error-Free, Synchronisation Error or No Synchronisation Error, Substitution Error or No Substitution Error, Insertion Error or No Insertion Error, and lastly Deletion Error or No Deletion Error. These are further explained in the next section.

\section{Error Category-Based Channel Models Analysis}
\label{sec:AnaChan}

\subsection{Error Category 1: Error or Error-Free}
\begin{figure}[ht]
\centering
\begin{subfigure}{\columnwidth}
		\centering
		\includegraphics[width=\linewidth]{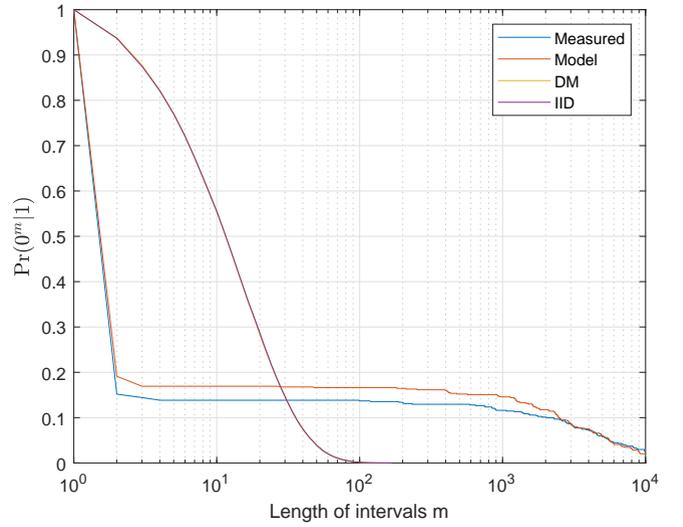}
        \caption{Error-free runs distribution for error or no error partitioning at low SNR data transmission\label{fig:ENEEFRLS}}
	\end{subfigure}
\begin{subfigure}{\columnwidth}
		\centering
		\includegraphics[width=\linewidth]{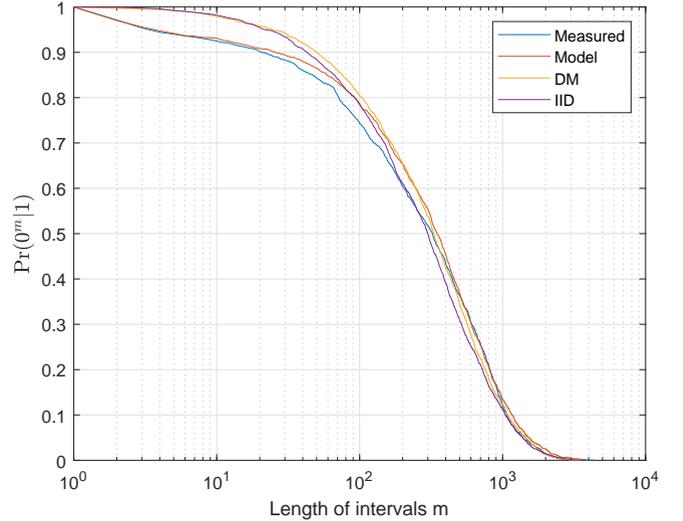}
        \caption{Error-free runs distribution for error or no error partitioning at high SNR data transmission\label{fig:ENEEFRHS}}
	\end{subfigure}
\caption[]{Error-free run distributions of Measured VLC data, simulated Model data, simulated DM model data and IID data when partitioning error sequence according to Error or no Error.}
\label{fig:ENEEFR}
	\end{figure}

Firstly, the synchronisation error sequence could be converted into a binary error sequence by looking for either an error or no error. In this case, all errors encountered (insertion, deletion, and substitution) in the synchronisation error sequence will be classified as an error and produce a $1$ in the binary error sequence, whereas no error is a perfect transmission and produces a $0$ in the binary error sequence. Using the example synchronisation error sequence from the previous section will produce \textbf{0,0,1,0,0,0,0,1,0,0,0,0,1,0,0,0,0,0,1,0} as the binary error sequence for this case. Figure \ref{fig:ENEEFR} shows the error-free run distributions for the measured VLC system data (Measured), the corresponding Fritchman models simulated data (Model), the DM model simulated data, and lastly an IID sequence. 

It is evident from Figure \ref{fig:ENEEFRLS}, which shows communication at low SNR, that the observed channel (Measured) contains a significant amount of memory as it deviates substantially from the IID plot which is by definition memoryless. The Model plot is almost identical to the measured data, which shows the Fritchman model and the parameters generated accurately depict the real channel. Additionally, the DM plot closely follows the IID plot which reiterates the idea that there is no memory between errors within the DM synchronisation model.
Communication at higher SNR, shown in Figure \ref{fig:ENEEFRHS}, shows that all the plots are similar, especially after run lengths of around 100 consecutive error-free transmissions. This follows intuition as there are less errors encountered at higher SNR values. Thus, the few errors produced during transmission would likely be more sporadic.

\begin{figure}[ht]
\centering
\begin{subfigure}{\columnwidth}
		\centering
		\includegraphics[width=\linewidth]{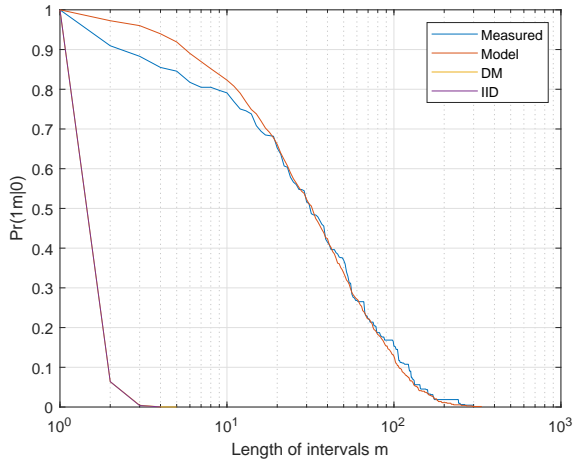}
        \caption{Error runs distribution for error or no error partitioning at low SNR data transmission}
        \label{fig:ENEERLS}
	\end{subfigure}
    
\begin{subfigure}{\columnwidth}
		\centering
		\includegraphics[width=\linewidth]{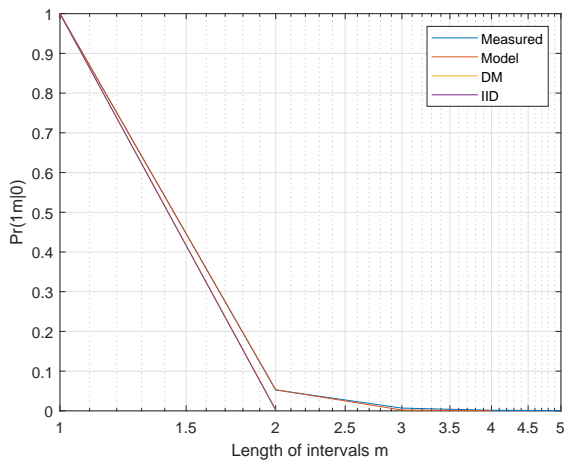}
        \caption{Error runs distribution for error or no error partitioning at high SNR data transmission}
        \label{fig:ENEERHS}
	\end{subfigure}
\caption[]{Error run distributions of measured VLC data, simulated model data, simulated Davey-Mackay model data and IID data when partitioning sequence according to Error or no Error.}
\label{fig:ENEER}
	\end{figure}

Looking at Figure \ref{fig:ENEER}, which shows the error-run distribution of the same models and parameters as above, similar trends are observed. In both the low and high SNR scenarios, the DM channel data closely follows that of the IID, while the Measured and Model data are highly correlated but deviate from the IID. The deviation is more significant in the case of the low SNR, shown in Figure \ref{fig:ENEERLS}. While there is still a difference between the measured data and IID in the high SNR, shown in Figure \ref{fig:ENEERHS}, it is almost negligible. We once again confirm the accuracy of the generated model as it accurately fits the measured data. It is also shown in the low SNR case that once an error is experienced, it is likely to cause another error. Variable cluster sizes are seen, sometimes exceeding over 100 consecutive erroneous digits. For the IID and DM data, the cluster of consecutive errors rarely exceeds 5 bits or symbols for the low SNR case. This is because there is no form of memory, and it is highly unlikely to see many consecutive errors.

\subsection{Error Category 2: Synchronisation Error or No Synchronisation Error}
\begin{figure}[ht]
\centering
\begin{subfigure}{\columnwidth}
		\centering
		\includegraphics[width=\linewidth]{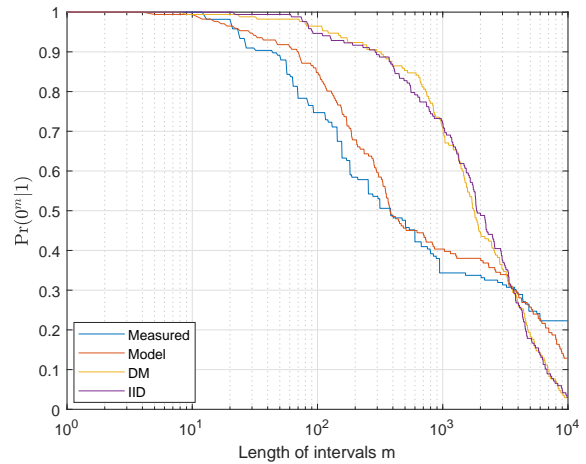}
        \caption{Error-free runs distribution for synchronisation error or no synchronisation error partitioning at low SNR data transmission}
        \label{fig:SNSEFRLS}
	\end{subfigure}
    
\begin{subfigure}{\columnwidth}
		\centering
		\includegraphics[width=\linewidth]{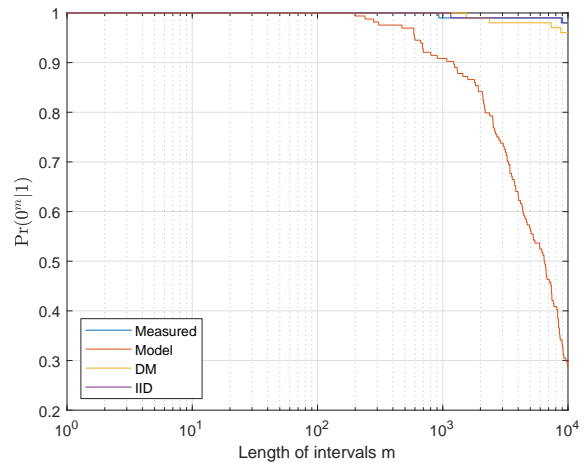}
        \caption{Error-free runs distribution for synchronisation error or no synchronisation error partitioning at high SNR data transmission}
        \label{fig:SNSEFRHS}
	\end{subfigure}

\caption[]{Error-free run distributions of measured VLC data, simulated model data, simulated Davey-Mackay model data and IID data when partitioning sequence according to synchronisation error or no synchronisation error.}
	\label{fig:SNSEFR}
	\end{figure}

Next, only synchronisation errors are isolated. In this case, a transmission and substitution error will produce a $0$ in the binary error sequence stream, whereas an insertion or deletion produces a $1$. Using the example synchronisation error sequence in Section \ref{sec:SetupChan} will produce \textbf{0,0,0,0,0,0,0,1,0,0,0,0,1,0,0,0,0,0,0,1} as the binary error sequence for this case. The error-free run distribution for the four plots with this partitioning of errors is shown in Figure \ref{fig:SNSEFR}. It is evident in the low SNR case, shown in Figure \ref{fig:SNSEFRLS}, that the DM model follows an IID trajectory quite closely while the Measured data, and subsequently the Fritchman Model data, deviates from it. This, again, indicates memory between symbols of correct transmission and synchronisation errors. The Model plot closely follows the Measured data, showing that the model created using the above process adequately describes our system characteristics. The plots in Figure \ref{fig:SNSEFRHS} are inconclusive as there were not enough synchronisation errors present at high SNR values to be accurately modeled and simulated.

\begin{figure}[th]
\centering
\includegraphics[width=\columnwidth]{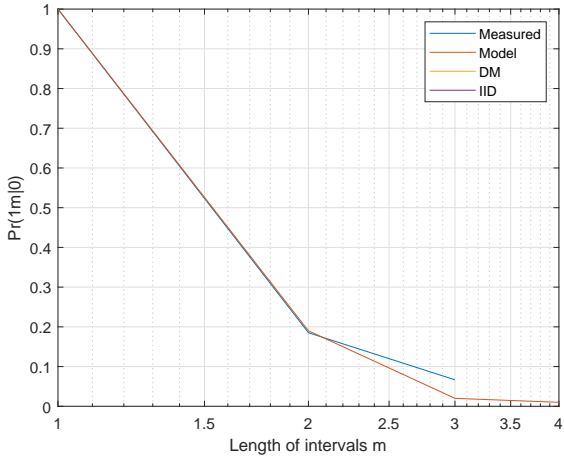}
\caption[]{Error run distributions of measured VLC data and simulated model data when partitioning sequence according to synchronisation error or no synchronisation error at low SNR data transmission.}
\label{fig:SNSER}
\end{figure}

Figure \ref{fig:SNSER} shows the error run distributions for the Measured VLC system data and the corresponding models simulated data for the synchronisation no synchronisation error partitioning at low SNR. It can be seen that there are at most two consecutive synchronisation errors for this system, and the occurrence of synchronisation errors, in general, are quite low. For this reason, the plots of the IID and DM channel are not visible, as simulating these channels with such low error probability allowed for at most a single synchronisation error between error-free runs. However, it is worth noting that while this system may not have substantial synchronisation errors, the procedure and methodology used can still be applied for harsher channels where more severe synchronisation errors do exist.

\subsection{Error Category 3: Substitution Error or No Substitution Error}
\begin{figure}[ht]
\centering
\begin{subfigure}{\columnwidth}
		\centering
		\includegraphics[width=\linewidth]{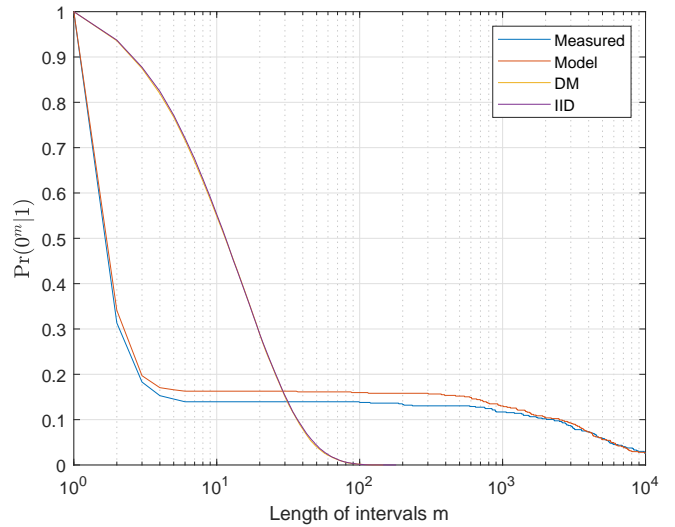}
        \caption{Error-free runs distribution for substitution error or no substitution error partitioning at low SNR data transmission}
        \label{fig:SuNSuEFRLS}
	\end{subfigure}
    
\begin{subfigure}{\columnwidth}
		\centering
		\includegraphics[width=\linewidth]{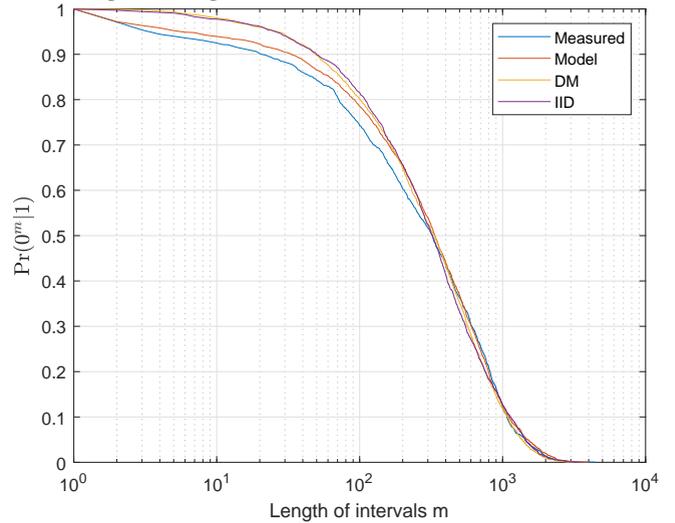}
        \caption{Error-free runs distribution for substitution error or no substitution error partitioning at high SNR data transmission}
        \label{fig:SuNSuEFRHS}
	\end{subfigure}

\caption[]{Error-free run distributions of measured VLC data, simulated model data, simulated Davey-Mackay model data and IID data when partitioning sequence according to substitution error or no substitution error.}
\label{fig:SuNSuEFR}
	\end{figure}

 In this error category, we focus on how substitution errors affect the error-free and error run distributions. In this scenario, a transmission and synchronisation error will produce a $0$ in the binary error sequence stream, whereas a substitution error produces a $1$. Using the example synchronisation error sequence from Section \ref{sec:SetupChan} will produce \textbf{0,0,1,0,0,0,0,0,0,0,0,0,0,0,0,0,0,0,1,0} as the binary error sequence for this case. The error-free run distribution for the various plots with this partitioning of errors is shown in Figure \ref{fig:SuNSuEFR}, where \ref{fig:SuNSuEFRLS} and \ref{fig:SuNSuEFRHS} show the plots for low SNR and high SNR communication respectively. These error-free and error run distribution plots are almost identical to the plots shown in Figure \ref{fig:ENEEFR} and Figure \ref{fig:ENEER}. This indicates that the most common type of error encountered in the system are substitution errors and a similar insight into them naturally follows.

	\begin{figure}[ht]
	\centering
\begin{subfigure}{\columnwidth}
		\centering
		\includegraphics[width=\linewidth]{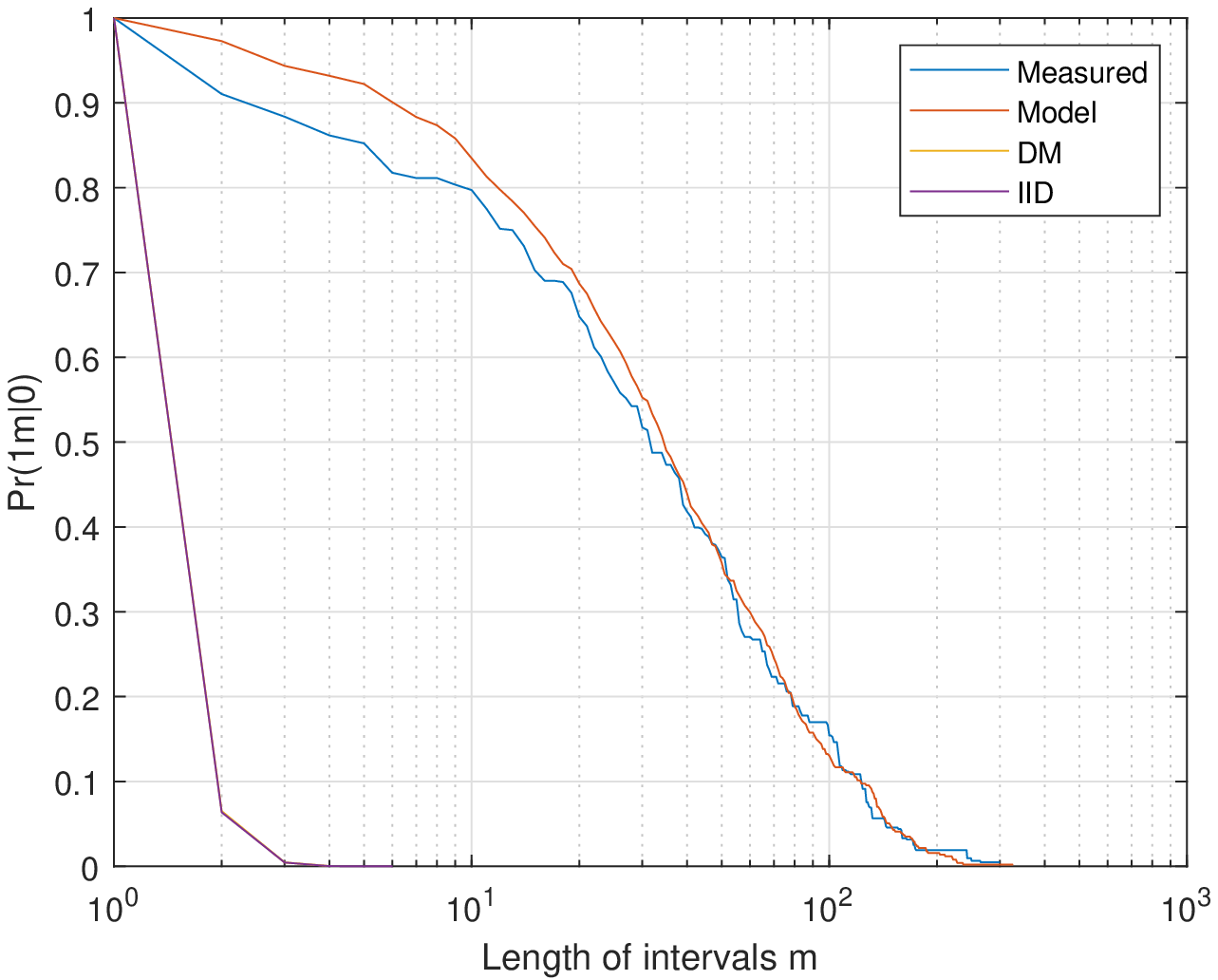}
        \caption{Error runs distribution for substitution error or no substitution error partitioning at low SNR data transmission}
        \label{fig:SuNSuERLS}
	\end{subfigure}
    
\begin{subfigure}{\columnwidth}
		\centering
		\includegraphics[width=\linewidth]{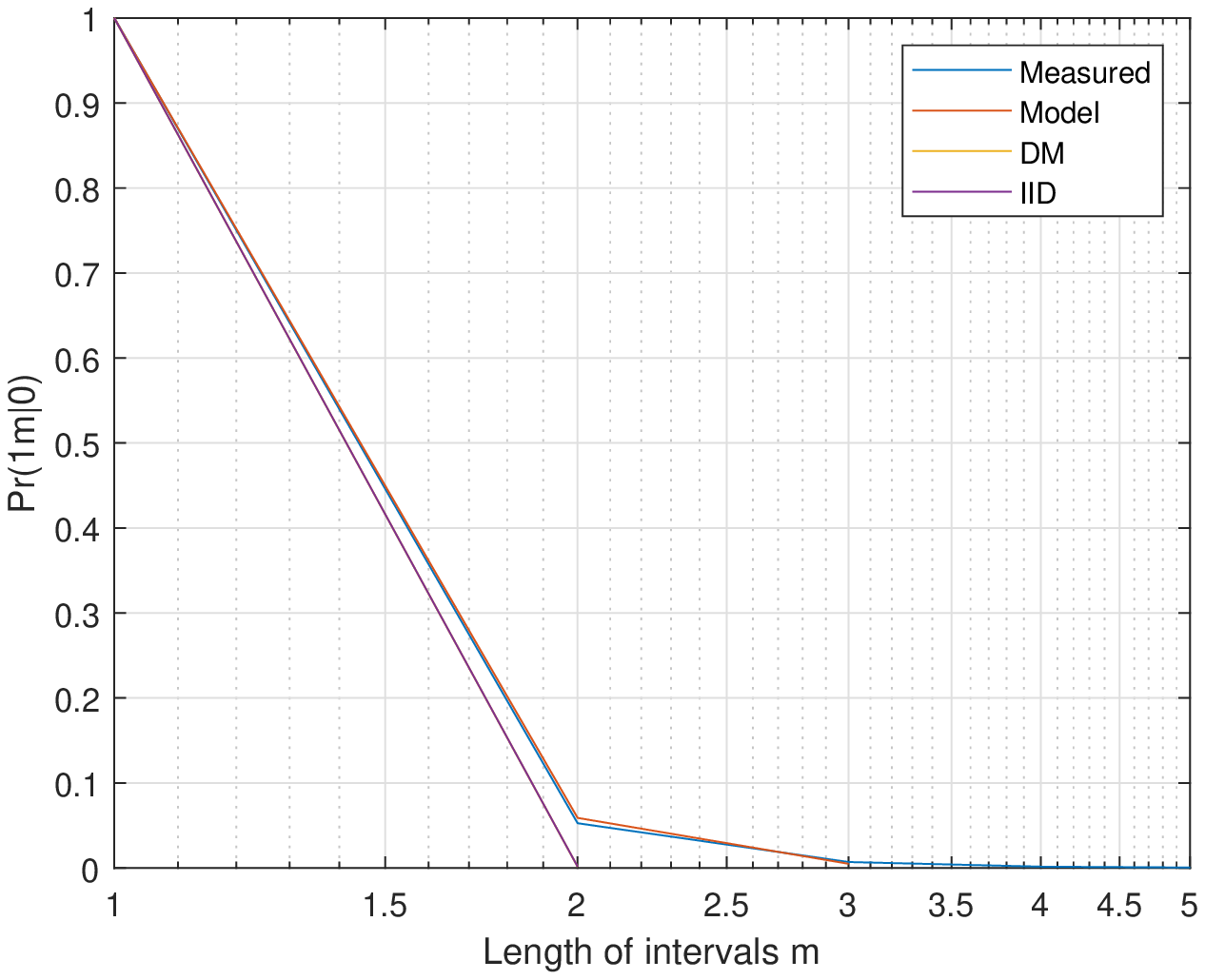}
        \caption{Error runs distribution for substitution error or no substitution error partitioning at high SNR data transmission}
        \label{fig:SuNSuERHS}
	\end{subfigure}

\caption[]{Error run distributions of measured VLC data, simulated model data, simulated Davey-Mackay model data and IID data when partitioning sequence according to substitution error or no substitution error.}
\label{fig:SuNSuER}
	\end{figure}

\subsection{Error Category 4: Insertion Error or No Insertion Error }
\begin{figure}[ht]
	\centering
\begin{subfigure}{\columnwidth}
		\centering
		\includegraphics[width=\linewidth]{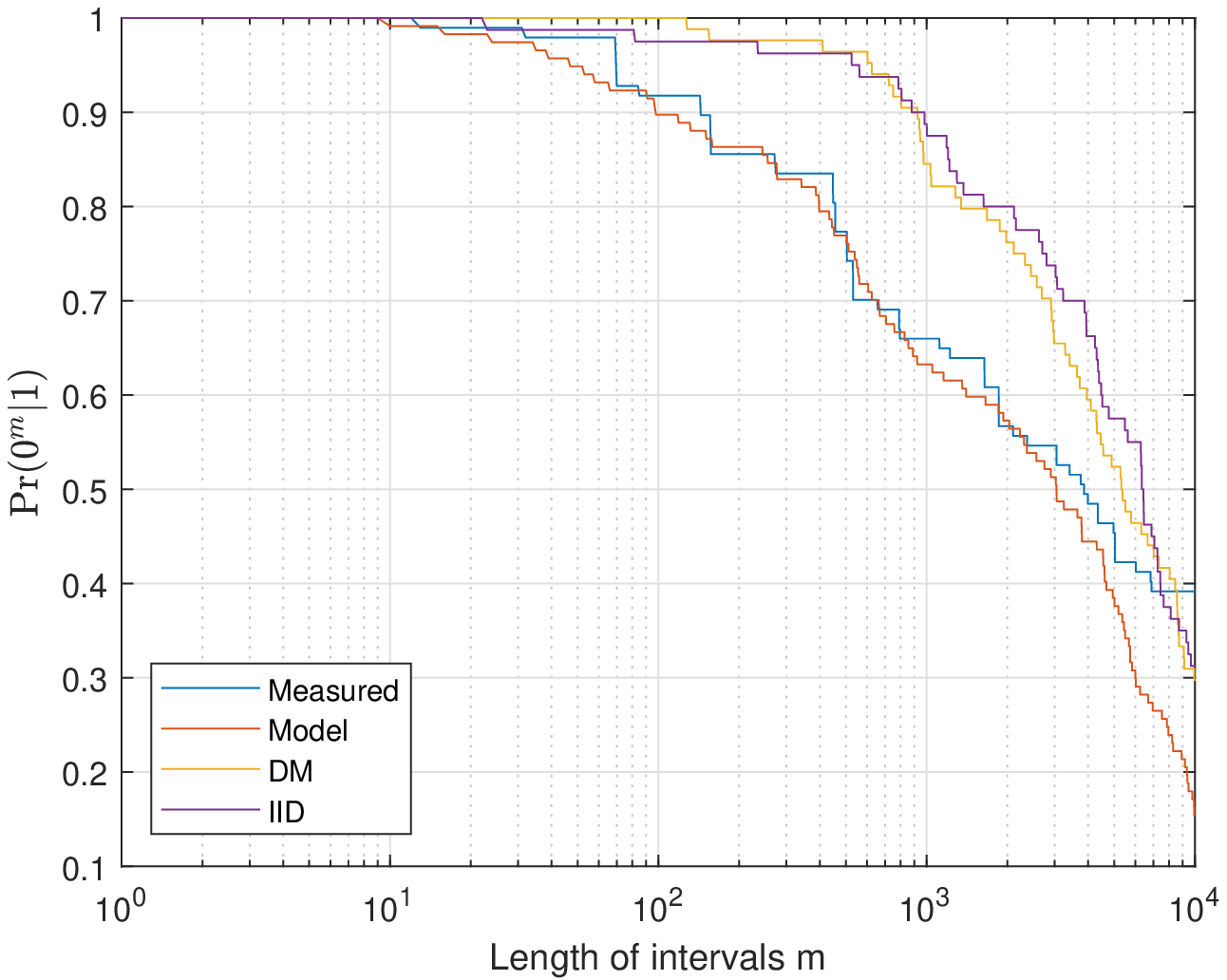}
        \caption{Error-free runs distribution for insertion error or no insertion error partitioning at low SNR data transmission}
        \label{fig:InIEFRLS}
	\end{subfigure}
    
\begin{subfigure}{\columnwidth}
		\centering
		\includegraphics[width=\linewidth]{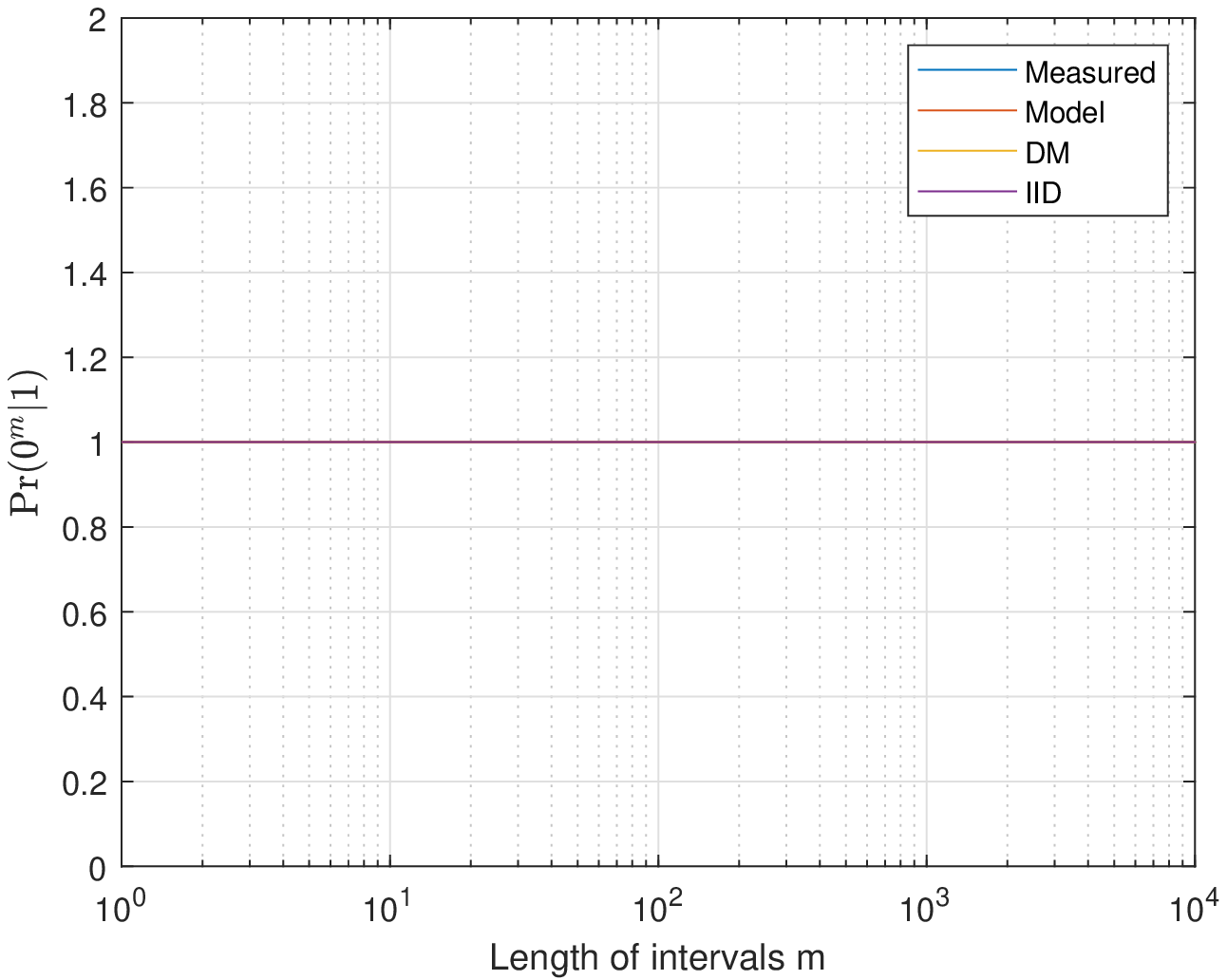}
        \caption{Error-free runs distribution for insertion error or no insertion error partitioning at high SNR data transmission}
        \label{fig:InIEFRHS}
	\end{subfigure}
	
\caption[]{Error run distributions of measured VLC data, simulated model data, simulated Davey-Mackay model data and IID data when partitioning sequence according to insertion error or no insertion error.}
\label{fig:InIEFR}
	\end{figure}

\begin{figure}[th]
\centering
\includegraphics[width=\columnwidth]{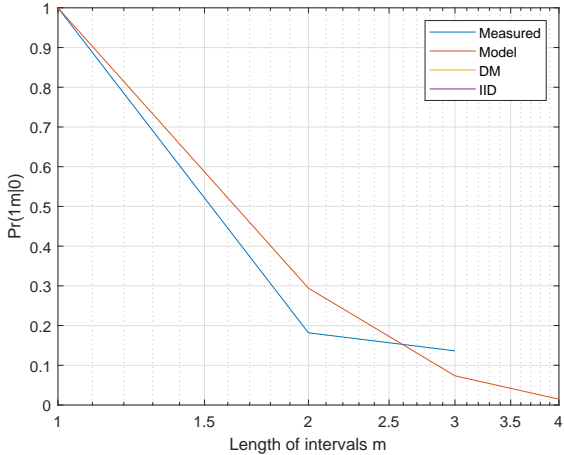}
\caption[]{Error run distributions of measured VLC data and simulated model data when partitioning sequence according to insertion error or no insertion error at low SNR data transmission.}
\label{fig:InIER}
\end{figure}

The following case is added for completeness of the analysis as the previous error categories already indicated that there is a low probability of producing synchronisation errors in the system, and subsequently insertion errors. For the case where partitioning is done according to insertion or no insertion error, the plots for the error-free runs and error runs are shown in Figure \ref{fig:InIEFR} and Figure \ref{fig:InIER}, where each sub figure shows low and high SNR data transmission. Only transmission for low SNR is shown for the error run distributions as there are too few insertions that occur at sufficiently high SNR communication. Since our interest for this error category is in the analysis of insertion errors, an insertion will produce a $1$ in the binary error sequence, while a transmission, substitution and deletion will all be analysed as no error and produce a $0$. For the given synchronisation error sequence this will produce a binary error sequence of \textbf{0,0,0,0,0,0,0,1,0,0,0,0,0,0,0,0,0,0,0,0}.

\subsection{Error Category 5: Deletion Error or No Deletion Error}
\begin{figure}[ht]
	\centering
\begin{subfigure}{\columnwidth}
		\centering
		\includegraphics[width=\linewidth]{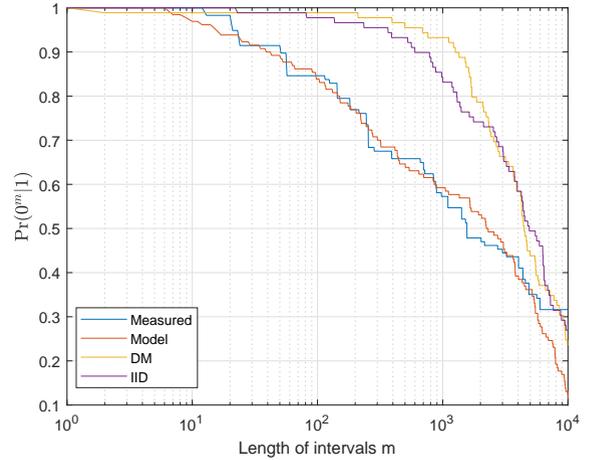}
        \caption{Error-free runs distribution for deletion error or no deletion error partitioning at low SNR data transmission}
        \label{fig:DnDEFRLS}
	\end{subfigure}
    
\begin{subfigure}{\columnwidth}
		\centering
		\includegraphics[width=\linewidth]{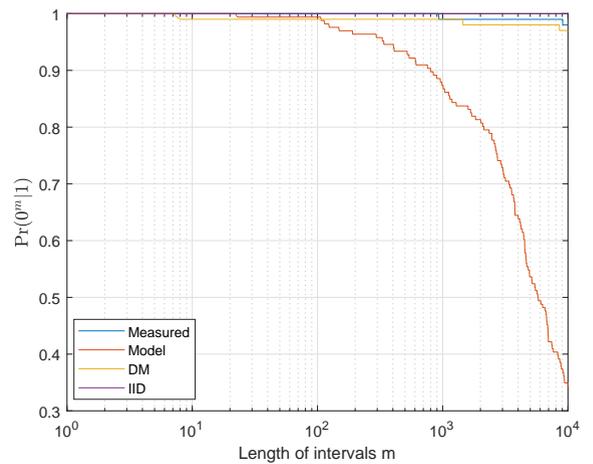}
        \caption{Error-free runs distribution for deletion error or no deletion error partitioning at high SNR data transmission}
        \label{fig:DnDEFRHS}
	\end{subfigure}
	
\caption[]{Error run distributions of measured VLC data, simulated model data, simulated Davey-Mackay model data and IID data when partitioning sequence according to deletion error or no deletion error.}
\label{fig:DnDEFR}
	\end{figure}

\begin{figure}[th]
\centering
\includegraphics[width=\columnwidth]{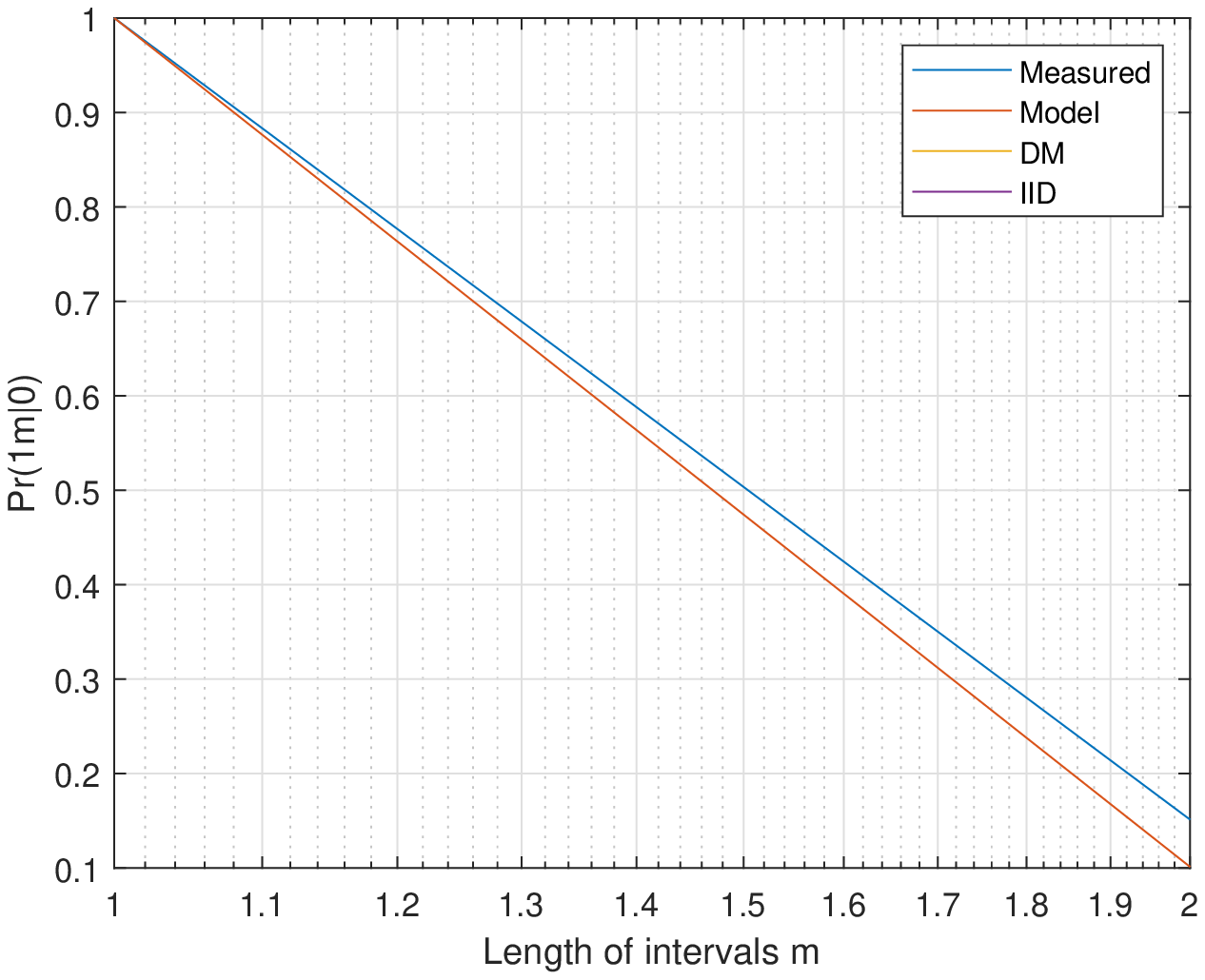}
\caption[]{Error run distributions of measured VLC data and simulated model data when partitioning sequence according to deletion error or no deletion error at low SNR data transmission.}
\label{fig:DnDER}
\end{figure}

Once again, the following case is added for completeness for the reasons explained above. For this case, where partitioning is done according to a deletion or no deletion error, the binary error sequence obtained from the example synchronisation error sequence is \textbf{0,0,0,0,0,0,0,0,0,0,0,0,1,0,0,0,0,0,0,0}. To obtain this sequence, a $1$ is placed in the binary error sequence when a deletion occurs in the synchronisation error sequence. All other likely states produce a $0$ in the binary error sequence. The plots for the error-free runs and error runs are shown in Figure  \ref{fig:DnDEFR} and Figure \ref{fig:DnDER}. Again, only the error-free runs have an associated low and high SNR communication. The error run distribution only indicates the low SNR communication as too few deletion errors are produced at high SNR communication. These plots once again are almost the same as the plots obtained for the cases of synchronisation error or no synchronisation error, and insertion error or no insertion error. As such, the analysis and observations would follow similar explanations.

\subsection{Chi Squared and MSE Analysis of Plots}
Table \ref{table:chimse} outlines the various Chi Squared and MSE values obtained for various plots under different communication parameters. While these values vary depending on the bin widths used (this may need to be optimised for a true representation), they still provide a good general indicator of how similar or contrasting the plots are to each other. In particular, it is noticed that at low SNR communication, the Model data and Measured data are generally in agreement with each other while significantly differing from the DM and IID channel plots. It is also evident that at these low SNR communication parameters, the DM plots are quite similar to an IID distribution. At high SNR communication, all the plots start to converge and we see a decrease in the Chi-Square and MSE values for these parameters. Again, we refer the reader to Appendix \ref{app:metricmethod} for a discussion on these metrics as well as the method implemented for the calculation of each statistic.

\begin{table*}\centering
\caption{Values obtained for Chi-Squared and MSE analysis for various parameters}
\label{table:chimse}
\resizebox{\textwidth}{!}{
\begin{tabular}{lllllllcllllll}\toprule
& \multicolumn{6}{c}{\textbf{Low SNR}}  && \multicolumn{6}{c}{\textbf{High SNR}}\\
\cmidrule{2-7} \cmidrule{9-14}

& \multicolumn{2}{c}{Bin \ Width = 1} & \multicolumn{2}{c}{Bin \ Width = 5} & \multicolumn{2}{c}{Bin \ Width = 10} && \multicolumn{2}{c}{Bin \ Width = 1} & \multicolumn{2}{c}{Bin \ Width = 5} & \multicolumn{2}{c}{Bin \ Width = 10}\\
\cmidrule{2-3} \cmidrule{6-7} \cmidrule{11-12} \\
& $\chi^2$ & $MSE$ & $\chi^2$ & $MSE$ & $\chi^2$ & $MSE$ && $\chi^2$ & $MSE$ & $\chi^2$ & $MSE$ & $\chi^2$ & $MSE$\\

\midrule
\textbf{Error No Error Plots}\\
IID vs Measured
& \makecell[l]{7202 \\ (p-value =0)\\ df = 35\\ Verdict: Reject $H_N$}
& \makecell[l]{8113 \\ $k=36$}
& \makecell[l]{1303 \\ (p-value =1.284e-271)\\ df = 12\\ Verdict: Reject $H_N$}
& \makecell[l]{14802 \\ $k=13$}
& \makecell[l]{550 \\ (p-value =1.485e-115)\\ df = 6\\ Verdict: Reject $H_N$}
& \makecell[l]{14977 \\ $k=7$}

&& \makecell[l]{6 \\ (p-value =0.01669)\\ df = 1\\ Verdict: Accept $H_N$}
& \makecell[l]{6385 \\ $k=2$}
& \makecell[l]{857 \\ (p-value =6.584e-125)\\ df = 91\\ Verdict: Reject $H_N$}
& \makecell[l]{239 \\ $k=92$}
& \makecell[l]{525 \\ (p-value =1.867e-74)\\ df = 62\\ Verdict: Reject $H_N$}
& \makecell[l]{425 \\ $k=63$}\\
\cmidrule{2-14}

IID vs DM 
& \makecell[l]{210 \\ (p-value =3.711e-12)\\ df = 87\\ Verdict: Reject $H_N$}
& \makecell[l]{545 \\ $k=88$}
& \makecell[l]{37 \\ (p-value =0.03568)\\ df = 23\\ Verdict: Accept $H_N$}
& \makecell[l]{1495 \\ $k=24$}
& \makecell[l]{25 \\ (p-value =0.02672)\\ df = 13\\ Verdict: Accept $H_N$}
& \makecell[l]{2775 \\ $k=14$}

&& \makecell[l]{7 \\ (p-value =0.006668)\\ df = 1\\ Verdict: Reject $H_N$}
& \makecell[l]{8192 \\ $k=2$}
& \makecell[l]{223 \\ (p-value =3.908e-13)\\ df = 91\\ Verdict: Reject $H_N$}
& \makecell[l]{62 \\ $k=92$}
& \makecell[l]{171 \\ (p-value =3.807e-12)\\ df = 62\\ Verdict: Reject $H_N$}
& \makecell[l]{83 \\ $k=63$}
\\ \cmidrule{2-14}

Model vs Measured 
& \makecell[l]{9 \\ (p-value =0.009418)\\ df = 2\\ Verdict: Reject $H_N$}
& \makecell[l]{324 \\ $k=3$}
& \makecell[l]{3 \\ (p-value =0.07644)\\ df = 1\\ Verdict: Accept $H_N$}
& \makecell[l]{349 \\ $k=2$}
& \makecell[l]{3 \\ (p-value =0.07644)\\ df = 1\\ Verdict: Accept $H_N$}
& \makecell[l]{349 \\ $k=2$}

&& \makecell[l]{12 \\ (p-value =0.03232)\\ df = 5\\ Verdict: Accept $H_N$}
& \makecell[l]{2701 \\ $k=6$}
& \makecell[l]{107 \\ (p-value =1.876e-13)\\ df = 21\\ Verdict: Reject $H_N$}
& \makecell[l]{107 \\ $k=22$}
& \makecell[l]{324 \\ (p-value =1.87e-28)\\ df = 89\\ Verdict: Reject $H_N$}
& \makecell[l]{69 \\ $k=90$}
\\ \cmidrule{2-14}

DM vs Measured 
& \makecell[l]{7201 \\ (p-value =0)\\ df = 35\\ Verdict: Reject $H_N$}
& \makecell[l]{8114 \\ $k=36$}
& \makecell[l]{2554 \\ (p-value =0)\\ df = 12\\ Verdict: Reject $H_N$}
& \makecell[l]{15349 \\ $k=13$}
& \makecell[l]{543 \\ (p-value =4.795e-114)\\ df = 6\\ Verdict: Reject $H_N$}
& \makecell[l]{14877 \\ $k=7$}

&& \makecell[l]{505 \\ (p-value =8.492e-112)\\ df = 1\\ Verdict: Reject $H_N$}
& \makecell[l]{2197 \\ $k=2$}
& \makecell[l]{885 \\ (p-value =1.384e-123)\\ df = 105\\ Verdict: Reject $H_N$}
& \makecell[l]{176 \\ $k=106$}
& \makecell[l]{497 \\ (p-value =1.38e-65)\\ df = 70\\ Verdict: Reject $H_N$}
& \makecell[l]{331 \\ $k=71$}
\\

\midrule

\textbf{Substitution Error No Substitution Error Plots}\\
IID vs Measured

& \makecell[l]{4643 \\ (p-value =0)\\ df = 33\\ Verdict: Reject $H_N$}
& \makecell[l]{5444 \\ $k=34$}
& \makecell[l]{2060 \\ (p-value =0)\\ df = 12\\ Verdict: Reject $H_N$}
& \makecell[l]{15010 \\ $k=13$}
& \makecell[l]{536 \\ (p-value =1.32e-112)\\ df = 6\\ Verdict: Reject $H_N$}
& \makecell[l]{14384 \\ $k=7$}

&& \makecell[l]{667 \\ (p-value =1.763e-145)\\ df = 2\\ Verdict: Reject $H_N$}
& \makecell[l]{2916 \\ $k=3$}
& \makecell[l]{673 \\ (p-value =1.821e-91)\\ df = 87\\ Verdict: Reject $H_N$}
& \makecell[l]{217 \\ $k=88$}
& \makecell[l]{541 \\ (p-value =5.438e-70)\\ df = 79\\ Verdict: Reject $H_N$}
& \makecell[l]{292 \\ $k=80$}
\\ \cmidrule{2-14}

IID vs DM 
& \makecell[l]{162 \\ (p-value =2.049e-06)\\ df = 87\\ Verdict: Reject $H_N$}
& \makecell[l]{478 \\ $k=88$}
& \makecell[l]{38 \\ (p-value =0.02852)\\ df = 23\\ Verdict: Accept $H_N$}
& \makecell[l]{2140 \\ $k=24$}
& \makecell[l]{20 \\ (p-value =0.06438)\\ df = 12\\ Verdict: Accept $H_N$}
& \makecell[l]{1405 \\ $k=13$}

&& \makecell[l]{6 \\ (p-value =0.04369)\\ df = 2\\ Verdict: Accept $H_N$}
& \makecell[l]{656 \\ $k=3$}
& \makecell[l]{148 \\ (p-value =5.249e-05)\\ df = 87\\ Verdict: Reject $H_N$}
& \makecell[l]{63 \\ $k=88$}
& \makecell[l]{157 \\ (p-value =3.797e-07)\\ df = 79\\ Verdict: Reject $H_N$}
& \makecell[l]{38 \\ $k=80$}
\\ \cmidrule{2-14}

Model vs Measured 
& \makecell[l]{3 \\ (p-value =0.4543)\\ df = 3\\ Verdict: Accept $H_N$}
& \makecell[l]{132 \\ $k=4$}
& \makecell[l]{2 \\ (p-value =0.1519)\\ df = 1\\ Verdict: Accept $H_N$}
& \makecell[l]{213 \\ $k=2$}
& \makecell[l]{2 \\ (p-value =0.1519)\\ df = 1\\ Verdict: Accept $H_N$}
& \makecell[l]{213 \\ $k=2$}

&& \makecell[l]{30 \\ (p-value =1.588e-05)\\ df = 5\\ Verdict: Reject $H_N$}
& \makecell[l]{544 \\ $k=6$}
& \makecell[l]{232 \\ (p-value =4.5e-16)\\ df = 83\\ Verdict: Reject $H_N$}
& \makecell[l]{52 \\ $k=84$}
& \makecell[l]{258 \\ (p-value =2.053e-18)\\ df = 89\\ Verdict: Reject $H_N$}
& \makecell[l]{78 \\ $k=90$}
\\ \cmidrule{2-14}

DM vs Measured 
& \makecell[l]{4538 \\ (p-value =0)\\ df = 35\\ Verdict: Reject $H_N$}
& \makecell[l]{5124 \\ $k=36$}
& \makecell[l]{2794 \\ (p-value =0)\\ df = 13\\ Verdict: Reject $H_N$}
& \makecell[l]{13702 \\ $k=14$}
& \makecell[l]{535 \\ (p-value =2.357e-112)\\ df = 6\\ Verdict: Reject $H_N$}
& \makecell[l]{14325 \\ $k=7$}

&& \makecell[l]{0 \\ (p-value =0.6809)\\ df = 1\\ Verdict: Accept $H_N$}
& \makecell[l]{181 \\ $k=2$}
& \makecell[l]{725 \\ (p-value =1.13e-111)\\ df = 66\\ Verdict: Reject $H_N$}
& \makecell[l]{252 \\ $k=67$}
& \makecell[l]{561 \\ (p-value =1.277e-71)\\ df = 84\\ Verdict: Reject $H_N$}
& \makecell[l]{240 \\ $k=85$}
\\ \cmidrule{2-14}
\midrule

& \multicolumn{2}{c}{Bin \ Width = 100} & \multicolumn{2}{c}{Bin \ Width = 500} & \multicolumn{2}{c}{Bin \ Width = 1000} && \multicolumn{2}{c}{Bin \ Width = 100} & \multicolumn{2}{c}{Bin \ Width = 500} & \multicolumn{2}{c}{Bin \ Width = 1000}\\
\cmidrule{2-3} \cmidrule{6-7} \cmidrule{11-12} \\
& $\chi^2$ & $MSE$ & $\chi^2$ & $MSE$ & $\chi^2$ & $MSE$ && $\chi^2$ & $MSE$ & $\chi^2$ & $MSE$ & $\chi^2$ & $MSE$\\

\midrule
\textbf{Synchronisation Error No Synchronisation Error Plots}\\
IID vs Measured
& \makecell[l]{239 \\ (p-value =1.331e-52)\\ df = 2\\ Verdict: Reject $H_N$}
& \makecell[l]{1607 \\ $k=3$}
& \makecell[l]{186 \\ (p-value =1.626e-34)\\ df = 10\\ Verdict: Reject $H_N$}
& \makecell[l]{408 \\ $k=11$}
& \makecell[l]{187 \\ (p-value =5.082e-37)\\ df = 7\\ Verdict: Reject $H_N$}
& \makecell[l]{792 \\ $k=8$}

&& \makecell[l]{0 \\ (p-value =1)\\ df = 1\\ Verdict: Accept $H_N$}
& \makecell[l]{0 \\ $k=2$}
& \makecell[l]{0 \\ (p-value =1)\\ df = 1\\ Verdict: Accept $H_N$}
& \makecell[l]{0 \\ $k=2$}
& \makecell[l]{0 \\ (p-value =1)\\ df = 1\\ Verdict: Accept $H_N$}
& \makecell[l]{0 \\ $k=2$}
\\ \cmidrule{2-14}

IID vs DM 
& \makecell[l]{2 \\ (p-value =0.3949)\\ df = 2\\ Verdict: Accept $H_N$}
& \makecell[l]{7 \\ $k=3$}
& \makecell[l]{8 \\ (p-value =0.5859)\\ df = 10\\ Verdict: Accept $H_N$}
& \makecell[l]{11 \\ $k=11$}
& \makecell[l]{3 \\ (p-value =0.9043)\\ df = 7\\ Verdict: Accept $H_N$}
& \makecell[l]{9 \\ $k=8$}

&& \makecell[l]{0 \\ (p-value =1)\\ df = 1\\ Verdict: Accept $H_N$}
& \makecell[l]{0 \\ $k=2$}
& \makecell[l]{0 \\ (p-value =1)\\ df = 1\\ Verdict: Accept $H_N$}
& \makecell[l]{0 \\ $k=2$}
& \makecell[l]{0 \\ (p-value =1)\\ df = 1\\ Verdict: Accept $H_N$}
& \makecell[l]{0 \\ $k=2$}
\\ \cmidrule{2-14}

Model vs Measured 
& \makecell[l]{21 \\ (p-value =0.000893)\\ df = 5\\ Verdict: Reject $H_N$}
& \makecell[l]{72 \\ $k=6$}
& \makecell[l]{21 \\ (p-value =2.831e-05)\\ df = 2\\ Verdict: Reject $H_N$}
& \makecell[l]{109 \\ $k=3$}
& \makecell[l]{2 \\ (p-value =0.1328)\\ df = 1\\ Verdict: Accept $H_N$}
& \makecell[l]{85 \\ $k=2$}

&& \makecell[l]{13 \\ (p-value =0.0002457)\\ df = 1\\ Verdict: Reject $H_N$}
& \makecell[l]{968 \\ $k=2$}
& \makecell[l]{68 \\ (p-value =3.468e-11)\\ df = 9\\ Verdict: Reject $H_N$}
& \makecell[l]{88 \\ $k=10$}
& \makecell[l]{275 \\ (p-value =2.447e-53)\\ df = 10\\ Verdict: Reject $H_N$}
& \makecell[l]{559 \\ $k=11$}
\\ \cmidrule{2-14}

DM vs Measured 
& \makecell[l]{296 \\ (p-value =5.112e-65)\\ df = 2\\ Verdict: Reject $H_N$}
& \makecell[l]{1765 \\ $k=3$}
& \makecell[l]{215 \\ (p-value =5.45e-42)\\ df = 8\\ Verdict: Reject $H_N$}
& \makecell[l]{552 \\ $k=9$}
& \makecell[l]{284 \\ (p-value =6.242e-56)\\ df = 9\\ Verdict: Reject $H_N$}
& \makecell[l]{686 \\ $k=10$}

&& \makecell[l]{0 \\ (p-value =1)\\ df = 1\\ Verdict: Accept $H_N$}
& \makecell[l]{0 \\ $k=2$}
& \makecell[l]{0 \\ (p-value =1)\\ df = 1\\ Verdict: Accept $H_N$}
& \makecell[l]{0 \\ $k=2$}
& \makecell[l]{0 \\ (p-value =1)\\ df = 1\\ Verdict: Accept $H_N$}
& \makecell[l]{0 \\ $k=2$}
\\
\midrule

\textbf{Insertion Error No Insertion Error Plots}\\
IID vs Measured
& \makecell[l]{2 \\ (p-value =0.1931)\\ df = 1\\ Verdict: Accept $H_N$}
& \makecell[l]{72 \\ $k=2$}
& \makecell[l]{2 \\ (p-value =0.1931)\\ df = 1\\ Verdict: Accept $H_N$}
& \makecell[l]{72 \\ $k=2$}
& \makecell[l]{70 \\ (p-value =1.231e-13)\\ df = 5\\ Verdict: Reject $H_N$}
& \makecell[l]{109 \\ $k=6$}

&& \makecell[l]{0 \\ (p-value =1)\\ df = 1\\ Verdict: Accept $H_N$}
& \makecell[l]{0 \\ $k=2$}
& \makecell[l]{0 \\ (p-value =1)\\ df = 1\\ Verdict: Accept $H_N$}
& \makecell[l]{0 \\ $k=2$}
& \makecell[l]{0 \\ (p-value =1)\\ df = 1\\ Verdict: Accept $H_N$}
& \makecell[l]{0 \\ $k=2$}
\\ \cmidrule{2-14}

IID vs DM 
& \makecell[l]{0 \\ (p-value =0.7389)\\ df = 1\\ Verdict: Accept $H_N$}
& \makecell[l]{5 \\ $k=2$}
& \makecell[l]{0 \\ (p-value =0.7389)\\ df = 1\\ Verdict: Accept $H_N$}
& \makecell[l]{5 \\ $k=2$}
& \makecell[l]{5 \\ (p-value =0.3603)\\ df = 5\\ Verdict: Accept $H_N$}
& \makecell[l]{7 \\ $k=6$}

&& \makecell[l]{0 \\ (p-value =1)\\ df = 1\\ Verdict: Accept $H_N$}
& \makecell[l]{0 \\ $k=2$}
& \makecell[l]{0 \\ (p-value =1)\\ df = 1\\ Verdict: Accept $H_N$}
& \makecell[l]{0 \\ $k=2$}
& \makecell[l]{0 \\ (p-value =1)\\ df = 1\\ Verdict: Accept $H_N$}
& \makecell[l]{0 \\ $k=2$}
\\ \cmidrule{2-14}

Model vs Measured 
& \makecell[l]{3 \\ (p-value =0.08377)\\ df = 1\\ Verdict: Accept $H_N$}
& \makecell[l]{93 \\ $k=2$}
& \makecell[l]{3 \\ (p-value =0.2041)\\ df = 2\\ Verdict: Accept $H_N$}
& \makecell[l]{33 \\ $k=3$}
& \makecell[l]{17 \\ (p-value =0.01051)\\ df = 6\\ Verdict: Accept $H_N$}
& \makecell[l]{40 \\ $k=7$}

&& \makecell[l]{0 \\ (p-value =1)\\ df = 1\\ Verdict: Accept $H_N$}
& \makecell[l]{0 \\ $k=2$}
& \makecell[l]{0 \\ (p-value =1)\\ df = 1\\ Verdict: Accept $H_N$}
& \makecell[l]{0 \\ $k=2$}
& \makecell[l]{0 \\ (p-value =1)\\ df = 1\\ Verdict: Accept $H_N$}
& \makecell[l]{0 \\ $k=2$}
\\ \cmidrule{2-14}

DM vs Measured 
& \makecell[l]{1 \\ (p-value =0.3374)\\ df = 1\\ Verdict: Accept $H_N$}
& \makecell[l]{41 \\ $k=2$}
& \makecell[l]{1 \\ (p-value =0.3374)\\ df = 1\\ Verdict: Accept $H_N$}
& \makecell[l]{41 \\ $k=2$}
& \makecell[l]{39 \\ (p-value =6.034e-07)\\ df = 6\\ Verdict: Reject $H_N$}
& \makecell[l]{67 \\ $k=7$}

&& \makecell[l]{0 \\ (p-value =1)\\ df = 1\\ Verdict: Accept $H_N$}
& \makecell[l]{0 \\ $k=2$}
& \makecell[l]{0 \\ (p-value =1)\\ df = 1\\ Verdict: Accept $H_N$}
& \makecell[l]{0 \\ $k=2$}
& \makecell[l]{0 \\ (p-value =1)\\ df = 1\\ Verdict: Accept $H_N$}
& \makecell[l]{0 \\ $k=2$}
\\
\midrule


\textbf{Deletion Error No Deletion Error Plots}\\
IID vs Measured
& \makecell[l]{4 \\ (p-value =0.05495)\\ df = 1\\ Verdict: Accept $H_N$}
& \makecell[l]{181 \\ $k=2$}
& \makecell[l]{197 \\ (p-value =2.251e-42)\\ df = 3\\ Verdict: Reject $H_N$}
& \makecell[l]{354 \\ $k=4$}
& \makecell[l]{102 \\ (p-value =1.588e-20)\\ df = 5\\ Verdict: Reject $H_N$}
& \makecell[l]{250 \\ $k=6$}

&& \makecell[l]{0 \\ (p-value =1)\\ df = 1\\ Verdict: Accept $H_N$}
& \makecell[l]{0 \\ $k=2$}
& \makecell[l]{0 \\ (p-value =1)\\ df = 1\\ Verdict: Accept $H_N$}
& \makecell[l]{0 \\ $k=2$}
& \makecell[l]{0 \\ (p-value =1)\\ df = 1\\ Verdict: Accept $H_N$}
& \makecell[l]{0 \\ $k=2$}
\\ \cmidrule{2-14}

IID vs DM 
& \makecell[l]{0 \\ (p-value =1)\\ df = 1\\ Verdict: Accept $H_N$}
& \makecell[l]{0 \\ $k=2$}
& \makecell[l]{8 \\ (p-value =0.04146)\\ df = 3\\ Verdict: Accept $H_N$}
& \makecell[l]{43 \\ $k=4$}
& \makecell[l]{18 \\ (p-value =0.003366)\\ df = 5\\ Verdict: Reject $H_N$}
& \makecell[l]{28 \\ $k=6$}

&& \makecell[l]{0 \\ (p-value =1)\\ df = 1\\ Verdict: Accept $H_N$}
& \makecell[l]{0 \\ $k=2$}
& \makecell[l]{0 \\ (p-value =1)\\ df = 1\\ Verdict: Accept $H_N$}
& \makecell[l]{0 \\ $k=2$}
& \makecell[l]{0 \\ (p-value =1)\\ df = 1\\ Verdict: Accept $H_N$}
& \makecell[l]{0 \\ $k=2$}
\\ \cmidrule{2-14}

Model vs Measured 
& \makecell[l]{2 \\ (p-value =0.5285)\\ df = 3\\ Verdict: Accept $H_N$}
& \makecell[l]{34 \\ $k=4$}
& \makecell[l]{3 \\ (p-value =0.2087)\\ df = 2\\ Verdict: Accept $H_N$}
& \makecell[l]{42 \\ $k=3$}
& \makecell[l]{11 \\ (p-value =0.02842)\\ df = 4\\ Verdict: Accept $H_N$}
& \makecell[l]{22 \\ $k=5$}

&& \makecell[l]{12 \\ (p-value =0.0004242)\\ df = 1\\ Verdict: Reject $H_N$}
& \makecell[l]{882 \\ $k=2$}
& \makecell[l]{27 \\ (p-value =5.03e-06)\\ df = 3\\ Verdict: Reject $H_N$}
& \makecell[l]{124 \\ $k=4$}
& \makecell[l]{151 \\ (p-value =3.068e-29)\\ df = 7\\ Verdict: Reject $H_N$}
& \makecell[l]{510 \\ $k=8$}
\\ \cmidrule{2-14}

DM vs Measured 
& \makecell[l]{4 \\ (p-value =0.04229)\\ df = 1\\ Verdict: Accept $H_N$}
& \makecell[l]{200 \\ $k=2$}
& \makecell[l]{4 \\ (p-value =0.04229)\\ df = 1\\ Verdict: Accept $H_N$}
& \makecell[l]{200 \\ $k=2$}
& \makecell[l]{334 \\ (p-value =3.645e-69)\\ df = 6\\ Verdict: Reject $H_N$}
& \makecell[l]{294 \\ $k=7$}

&& \makecell[l]{0 \\ (p-value =1)\\ df = 1\\ Verdict: Accept $H_N$}
& \makecell[l]{0 \\ $k=2$}
& \makecell[l]{0 \\ (p-value =1)\\ df = 1\\ Verdict: Accept $H_N$}
& \makecell[l]{0 \\ $k=2$}
& \makecell[l]{0 \\ (p-value =1)\\ df = 1\\ Verdict: Accept $H_N$}
& \makecell[l]{0 \\ $k=2$}
\\

\bottomrule
\end{tabular}
}

\end{table*}

\section{A Novel Memory Synchronisation Channel Model}
\label{sec:Novel}
While the above setup and analysis is useful for modeling memory in synchronisation channels, it is not without limitations as only specific, unique errors may be analysed at a given instance. In this section, a new model is developed which can incorporate memory and multiple types of errors within the channel simultaneously, as opposed to converting the errors into a binary form first.

\begin{figure}[th]
\centering
\includegraphics[width=\columnwidth]{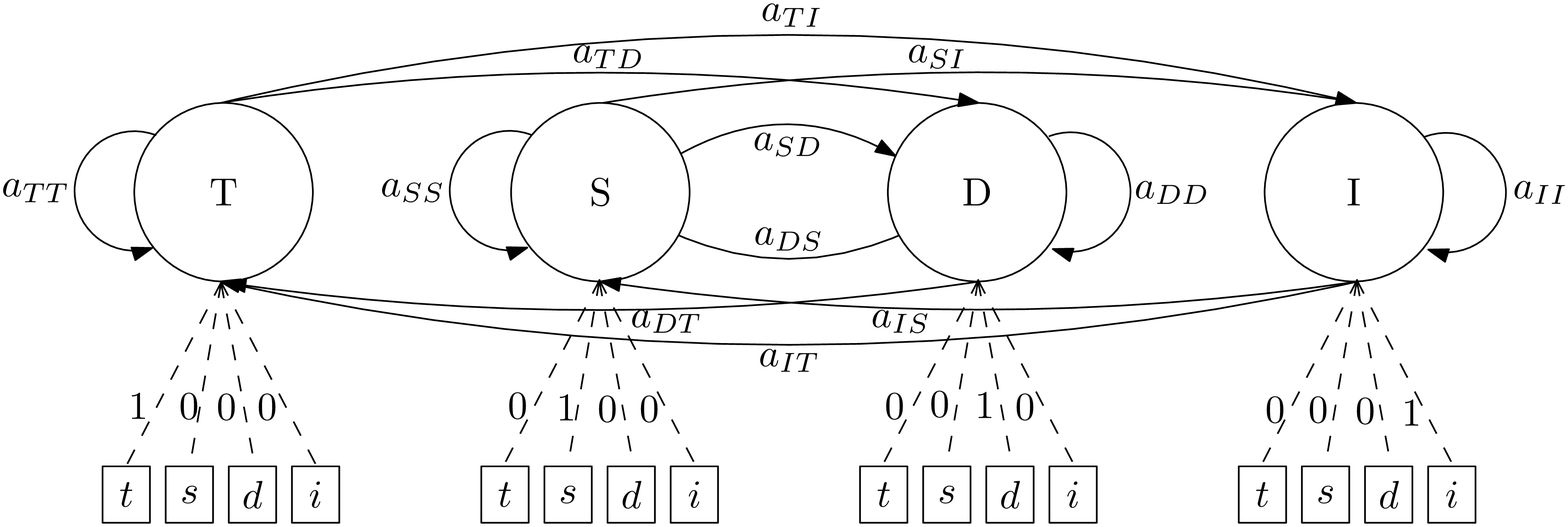}
\caption[Four state Markov model for IDS channel]{Four state Markov model for IDS channel}
\label{fig:newmod1}
\end{figure}

We again assume the receiver has full knowledge of the transmitted data and that each state emits a unique symbol $(t,s,d \text{ or } i)$ into the synchronisation error sequence to identify that particular state ie. that the states are completely visible. This reduces the model to a simple Markov chain. For this analysis, the synchronisation error sequence is used directly, where the method of producing the synchronisation error sequence is identical to that described in Section \ref{sec:SetupChan}. This sequence, along with the emission matrix, which is a diagonal of ones, will produce the transition matrix, $A$, for a four-state Markov model, which is illustrated in Figure \ref{fig:newmod1}, when run through the Baum-Welch algorithm. This process, while simple, produces accurate transition matrices for the memory synchronisation channel.

\subsection{Memory Synchronisation Channel Models obtained from real-world data}
\begin{figure}[th]
\centering
\includegraphics[width=\columnwidth]{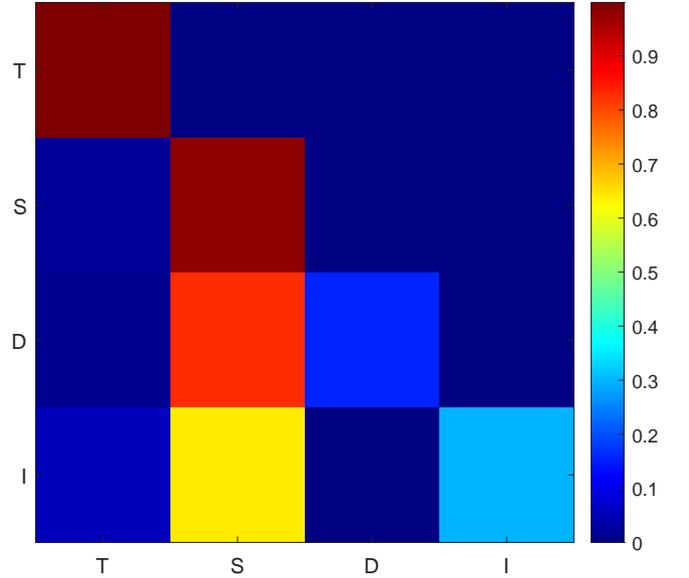}
\caption[Transition matrix heat map at low SNR communication]{Transition matrix heat map at low SNR communication}
\label{fig:HMlow}
\end{figure}

\begin{figure}[th]
\centering
\includegraphics[width=\columnwidth]{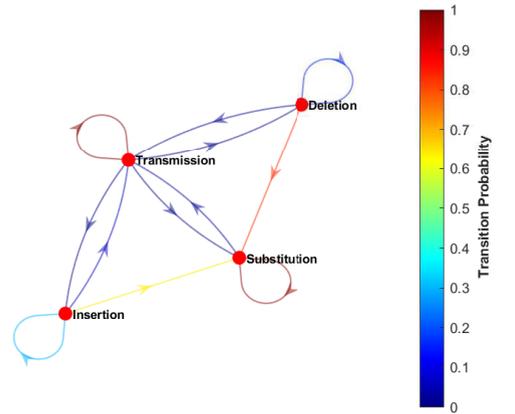}
\caption[Directed graph for proposed IDS memory model at low SNR communication]{Directed graph for proposed IDS memory model at low SNR communication}
\label{fig:DGlow}
\end{figure}

We once again apply this technique to the data from the VLC system in \cite{achari2021selfsynchronising} with the same parameters as before. Figure \ref{fig:HMlow} shows the transition matrix heatmap for low SNR communication in the VLC system. The actual values for the transition matrix are shown in Equation \eqref{eqn:lowTrans}. From this a directed graph which represents the Markov chain can be plot. This is illustrated in Figure \ref{fig:DGlow}. It is evident from this model that once a transition occurs, the channel is most likely going to remain in this state. The same thing occurs with substitution errors, as the self transition probability of this state is quite high. It is also worth noting that the probability of transitioning from a synchronisation error (either deletion or insertion) to a substitution error is also quite high, while transitioning from any error back to normal transmission is relatively small. This adequately illustrates the memory we notice between the three error states, as well as the long runs of error-free transmissions observed.

\begin{figure}[th]
\centering
\includegraphics[width=\columnwidth]{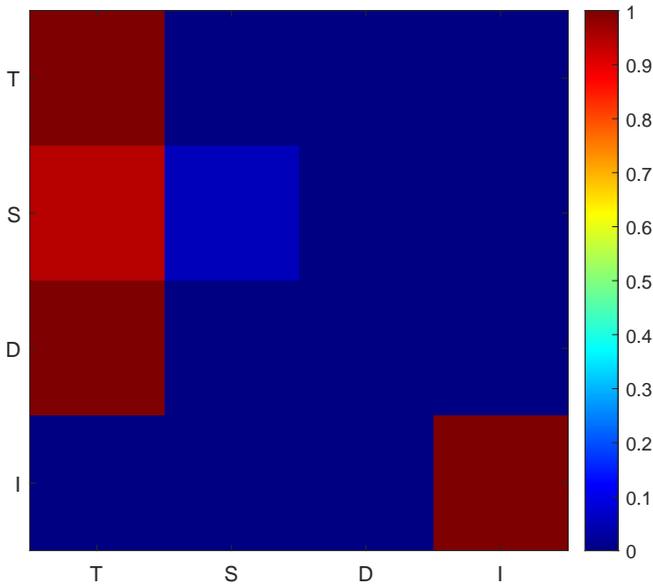}
\caption[Transition matrix heat map at high SNR communication]{Transition matrix heat map at high SNR communication}
\label{fig:HMhigh}
\end{figure}

\begin{figure}[th]
\centering
\includegraphics[width=\columnwidth]{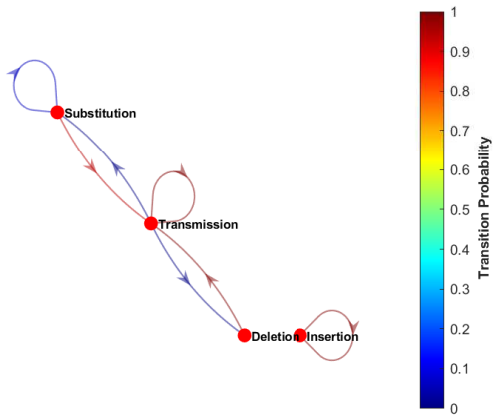}
\caption[Directed graph for proposed IDS memory model at high SNR communication]{Directed graph for proposed IDS memory model at high SNR communication}
\label{fig:DGhigh}
\end{figure}

The same analysis is performed for communication at higher SNR values, which yields the transition matrix heatmap shown in Figure \ref{fig:HMhigh}, and the directed graph shown in Figure \ref{fig:DGhigh}. Equation \eqref{eqn:highTrans} represents the transition matrix obtained for high SNR communication. It is evident from the high SNR communication analysis that once the system is in an error-free (transmission) state, it will more than likely continue in this state. Even if a substitution error occurs, the system quickly returns to the transmission state with very little probability of consecutive errors. It is even shown that as soon a single deletion occurs, the system still immediately returns to error free transmission. It is worth noting that since no insertions are observed at high SNR, we assume there are no transitions out of this state, hence the insertion state is isolated in Figure \ref{fig:DGhigh}.

\begin{equation}
\label{eqn:lowTrans}
A = 
\begin{bmatrix} 
0.9980 & 0.0020 & 1.0120-6 & 0\\
0.9421 & 0.0579 & 0 & 0\\
1 & 0 & 0 & 0\\
0 & 0 & 0 & 0\\
\end{bmatrix}
\quad
\end{equation}

\begin{equation}
\label{eqn:highTrans}
A = 
\begin{bmatrix} 
0.9986 & 0.0011 & 1.5346e-4 & 1.0898-4\\
0.0203 & 0.9797 & 0 & 0\\
0.0122 & 0.8293 & 0.1585 & 0\\
0.0571 & 0.6429 & 0 & 0.3\\
\end{bmatrix}
\quad
\end{equation}

\section{Conclusion}
\label{sec:Conclusion}
Various memory models and synchronisation error channels are discussed, but there is unfortunately no overlap which accounts for IDS channels that contain statistically dependent errors. Firstly,a novel technique to determine channel characteristics and model parameters is introduced which builds onto the idea of the Fritchman model while making use of the Levenshtein distance and different error categories.  The proposed channel models show a clear distinction in the error and error-free runs from the DM channel, which seeks to only model statistically independent synchronisation errors. Finally, a more encompassing model, which can be viewed as a Markov chain, that accounts for insertion, deletion, and substitution errors is described. The new method is more enveloping of practical, real world communication channels and is demonstrated by making use of data from a VLC system. This technique may even be applied to other channels in applications such as the barcoding of DNA sequences.

\bibliographystyle{IEEEtran}
\bibliography{ref}

\begin{IEEEbiography}[{\includegraphics[width=1in,height=1.25in,clip,keepaspectratio]{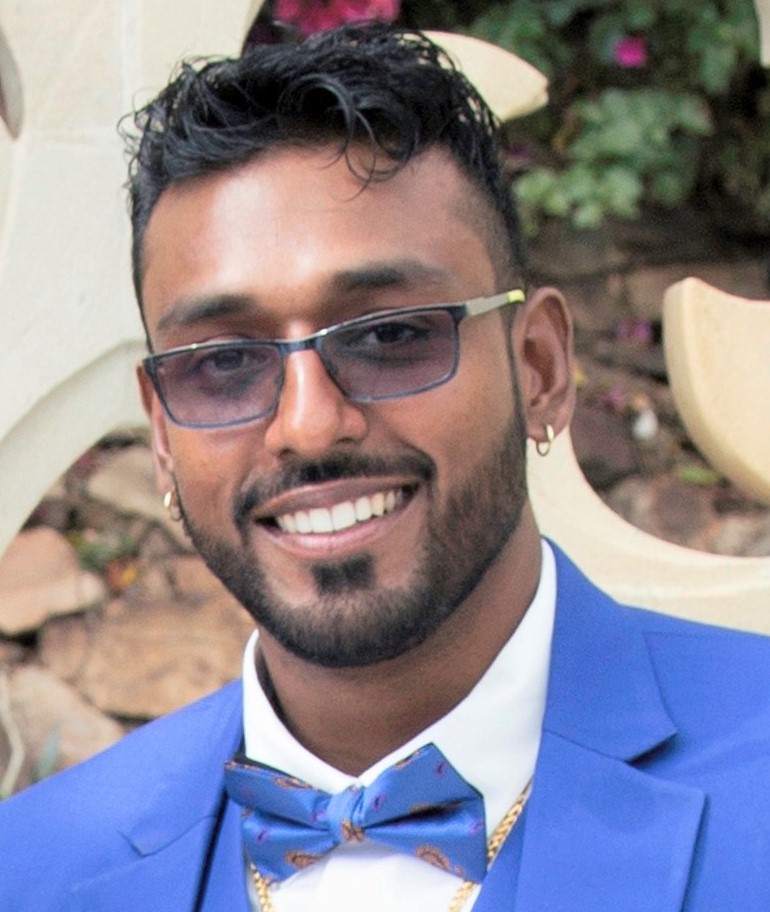}}]{Shamin Achari} received a  BSc degree in electrical and information engineering from the University of the Witwatersrand (WITS) in 2015. He is currently pursuing the Ph.D. degree in electrical engineering at the University of the Witwatersrand Johannesburg South Africa.
His MSc was based in the field of Visible Light Communication with special focus on error correction schemes for visible light systems which was subsequently upgraded to a Ph.D. which mainly focuses on channel modelling and methods of correcting synchronisation errors.
His research interests include visible light communications, machine learning and Internet of Things (IoT).
\end{IEEEbiography}

\begin{IEEEbiography}[{\includegraphics[width=1in,height=1.25in,clip,keepaspectratio]{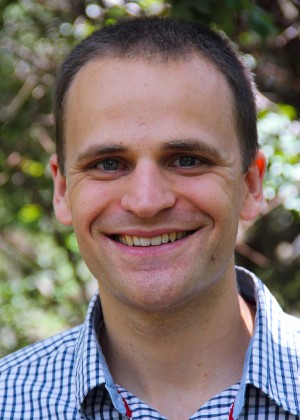}}]{Daniel G. Holmes} received his BSc Eng. in Electrical and Information Engineering from the University of the Witwatersrand, Johannesburg, in 2015. He was awarded the Adolf Goldsmith Memorial Fund Prize in 2015 by the Wits School of Electrical and Information Engineering in recognition of the significant progress made in his undergraduate degree. Daniel served as the chair of the SAIEE-IEEE Wits Student Branch during 2016 and 2017. In 2018 he graduated with an MSc Eng. in the field of visible light communication channel modelling at the University of the Witwatersrand.
\end{IEEEbiography}

\begin{IEEEbiography}[{\includegraphics[width=1in,height=1.25in,clip,keepaspectratio]{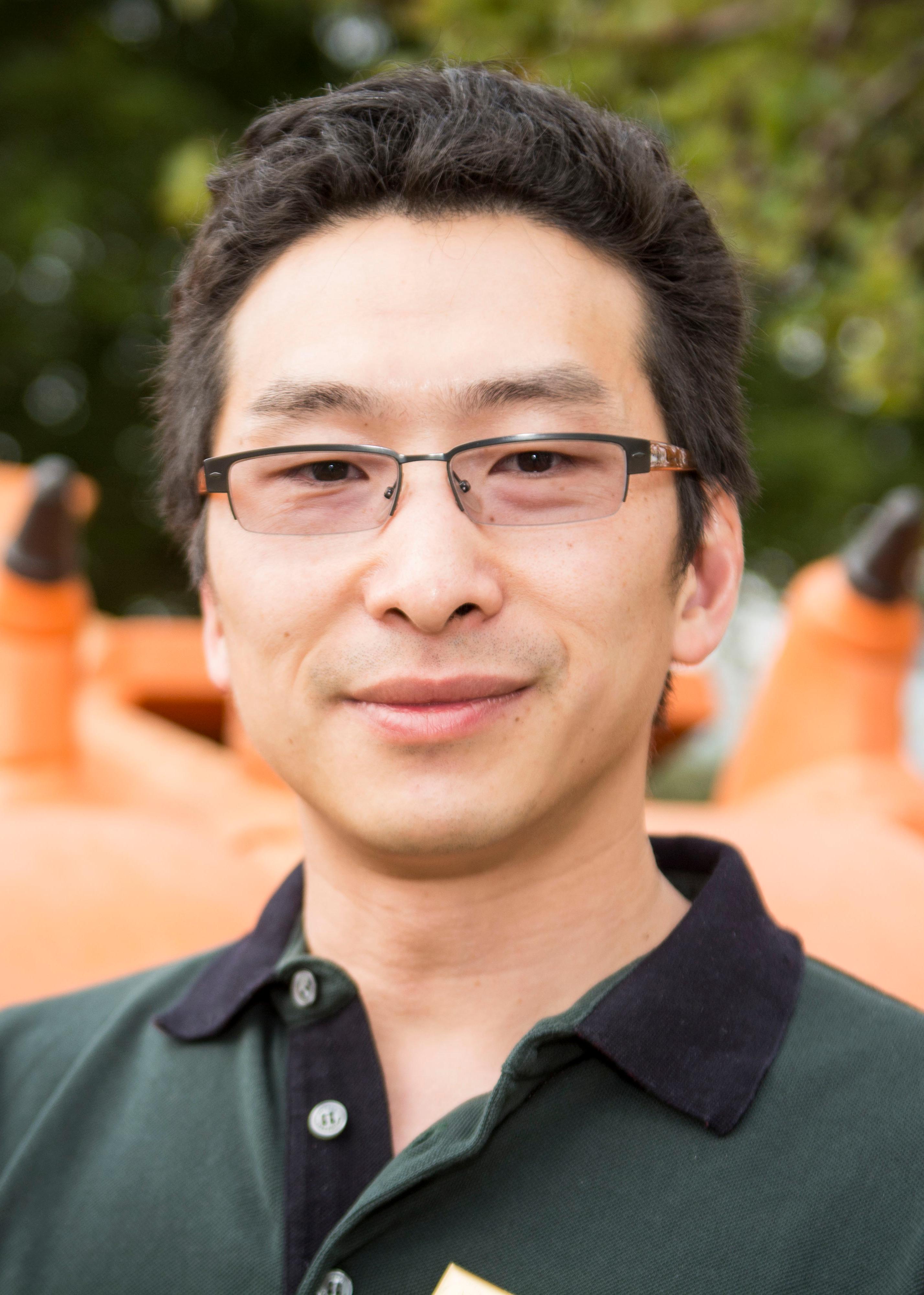}}]{Ling Cheng} (M'10--SM'15) received the degree B.Eng. Electronics and Information (cum laude) from Huazhong University of Science and Technology (HUST) in 1995, M. Ing. Electrical and Electronics (cum laude) in 2005, and D. Ing. Electrical and Electronics in 2011 from University of Johannesburg (UJ). His research interests are in Telecommunications and Artificial Intelligence. In 2010, he joined University of the Witwatersrand where he was promoted to Full Professor in 2019. He serves as the associate editor of three journals. He has published more than 100 research papers in journals and conference proceedings. He has been a visiting professor at five universities and the principal advisor for over forty full research post-graduate students. He was awarded the Chancellor’s medals in 2005, 2019 and the National Research Foundation rating in 2014. The IEEE ISPLC 2015 best student paper award was made to his Ph.D. student in Austin. He is a senior member of IEEE and the vice-chair of IEEE South African Information Theory Chapter. 

\end{IEEEbiography}

\appendix

\setcounter{figure}{0}
\setcounter{equation}{0}

\section{}

\subsection{Performance Metrics for Analysis and Methodology}\label{app:metricmethod}
\renewcommand{\theequation}{\thesubsection.\arabic{equation}}
\renewcommand{\thefigure}{\thesubsection.\arabic{figure}}

To compare and quantify the similarities, and consequently differences,between the various plots produced; the Chi Square ($\chi^2$) and Mean Squared Error (MSE) metrics are used. This Appendix outlines the background on the various analysis metrics used as well as the procedures and methods used when calculating these quantities.

\subsubsection{Chi-Squared ($\chi^2$) goodness of fit}
The Chi-Squared test is a non-parametric test that is used to determine if the observed data is significantly different from expected data and is calculated using Equation \eqref{eqn:chi} where $k$ is the number of data samples, $o_i$ is the observed data at sample $i$ and $e_i$ is the expected value of the $i$th sample  \cite{kaChi, zstatChi,gingrich1992introductory}. 

\begin{equation}
\label{eqn:chi}
\chi^2 = \sum_{i=1}^{k} \frac{(o_i-e_i)^2}{e_i}
\end{equation}

As this is a method used in hypothesis testing, two hypotheses first need to be constructed and later tested. The null hypothesis $H_N$ describes the situation where there is no significant difference in the distributions of the compared plots whereas the alternative hypothesis, $H_A$, describes the situation where there is a significant difference in distribution between the compared plots. For all the tests, a strict significance value of $0.01$ is used, but it is not uncommon to use 5\% or 10\% depending on the scenario. The degrees of freedom varies as a function of the bin size and the context of the plots.

\subsubsection{Mean Squared Error (MSE)}

The Mean Squared Error is a metric that is traditionally used to indicate how close a regression line is to a set of observed values \cite{freecodeMSE}. Here we will use the MSE, in a similar fashion to the $\chi^2$ metric, as an indicator of how similar the various error-free run plots are. The MSE equation is described by Equation \eqref{eqn:mse} where the variables are defined as before for the $\chi^2$ equation. Unlike the Chi-Squared metric, there is no way of determining a good fit with the MSE value alone and thus we will need to compare different MSE values from different plots against each other, where a larger value of the MSE indicates a more significant difference between the plots compared. 
 
\begin{equation}
\label{eqn:mse}
MSE = \frac{1}{k} \sum_{i=1}^{k}(o_i-e_i)^2
\end{equation}

\subsubsection{Error-free Run and Error Run Distributions}
Two of the most important metrics used to define discrete channel models are the error-free run distribution and the error run distribution. The error-free run distribution is defined as $Pr(0^m|1)$, which is the probability of receiving a stream of $m$ or more consecutive error-free transmissions following an error. An error run distribution is defined as $Pr(1^m|0)$ and describes the probability of receiving $m$ or more consecutive errors after an error-free transmission \cite{tranter2004principles}. 

In general, for the Fritchman model, the error-free runs and error runs are given by Equations \eqref{eqn:FrEFR} and \eqref{eqn:FrER} respectively, which describes these distribution events in terms of weighted exponentials \cite{tranter2004principles,fritchman1967binary}. Here, $\lambda_i$ represents the eigenvalues of $A_{gg}$ and $A{bb}$ where $A_{gg}$ and $A{bb}$ are the diagonal sub-matrices which form part of the transition matrix $A$ and $f_i$ is the corresponding transition probability from $a_{ij}$ \cite{fritchman1967binary,tranter2004principles}. For a single error state Fritchman model, Equation \eqref{eqn:FrEFR} is simplified to Equation \eqref{eqn:FrEFR2} \cite{tranter2004principles}.

\begin{equation}
\label{eqn:FrEFR}
Pr(0^m|1) = \sum_{i=1}^{k} f_i \lambda_i^{m-1}
\end{equation}

\begin{equation}
\label{eqn:FrER}
Pr(1^m|0) = \sum_{i=k+1}^{k} f_i \lambda_i^{m-1}
\end{equation}

\begin{equation}
\label{eqn:FrEFR2}
Pr(0^m|1) = \sum_{k=1}^{N-1} \frac{a_{Nk}(a_{kk})^m}{a_{kk}}
\end{equation}

\subsubsection{Procedure for $\chi^2$ and MSE Calculations }

For both the  $\chi^2$ and the MSE, the data from the error-free runs is segmented into different bins according to a predefined bin width. Bin widths of 1, 5 and 10 are chosen for Error vs Error-Free segmentation as well as Substitution Error vs No Substitution Error segmentation while bin widths of 100, 500 and 1000 are selected for the remaining segmentation plots. The rationale behind this is to ensure that the large counts condition is met. The large counts condition states that each category (in our case each bin) has an expected outcome of at least 5 \cite{kaChi}. This is done to ensure the criteria of the $\chi^2$ analysis is met as five or more occurrences in each expected bin satisfies the criteria for the central limit theorem which allows the distributions to be normally distributed in nature ie. there needs to be a large enough samples for the central limit theorem to be met as the $\chi^2$ statistic is based on a normal distribution \cite{zstatChi}. 
Additionally, the traditional definition of the error free runs, Pr($0^m$|1), which is the probability of receiving a stream of \textbf{at least}  $m$ consecutive error-free transmissions following an error is replaced by a stricter criteria in the $\chi^2$ and the MSE analysis where the bins are categorised by \textbf{exactly} m error-free transmissions following an error. In other words, if the bin width is set to 5, we would have categories or bins of 1 to 5 error-free transmissions, 6 to 10 error-free transmissions, 11 to 15 error-free transmissions and so on. Here the bin 1 to 5 error-free transmissions does not represent at least 1 to 5 error-free transmissions but rather how many times exactly 1,2,3,4 or 5 error-free transmissions occur. This is done to satisfy the criteria that $\sum_{i=1}^{k} r_i = 1$ where $r_i$ is the proportion or percentage of counts in each category \cite{gingrich1992introductory}. In other words all counts are independent and cannot attribute to multiple categories. Finally, a cutoff for the number of consecutive error free runs must be set as many of the longer chains of error-free transmissions become less frequent and often times the counts are less than five occurrences. Again, the rationale is to satisfy the large counts criteria so that the $\chi^2$ analysis is valid. As the longer runs of error-free transmissions are less likely to occur in the expected variable, the first bin that has a frequency or occurrence count less than 5 is selected as the cutoff value and all subsequent categories, including the cutoff bin, are grouped together as a single bin.
 Once the segmentation of data into bins is complete the $\chi^2$ and MSE statistics may now be calculated using Equation \eqref{eqn:chi} and Equation \eqref{eqn:mse} respectively. 

For the $\chi^2$ statistic, the p-value is then calculated using the degrees of freedom which corresponds to the number of bins less one. Once the p-value is obtained we can then compare this to the significance level originally chosen for the analysis. If the p-value is less than the significance level, the null hypothesis is rejected and consequently the alternate hypothesis is accepted \cite{kaChi, zstatChi, gingrich1992introductory}. If the p-value is greater than the significance level, it implies there is not enough evidence to suggest the null hypothesis is wrong and thus we accept it as true. It is worth noting that as the bin width increases, the degrees of freedom (categories) decreases. In some cases there is only one bin that contains all counts and the corresponding Chi Square and MSE values tell us if the total expected number of errors is in agreement to the total measured number of errors for the given parameters. Furthermore we note that more accurate values for these metrics will be obtained when using optimised bin widths for given parameters.

\EOD
\end{document}